\documentclass[journal]{IEEEtran}

\usepackage{amsmath,amssymb}
\usepackage{graphicx}
\graphicspath{{./}}
\usepackage{booktabs}
\usepackage{tabularx}
\usepackage{cite}
\usepackage{url}

\title{MAG-Net: Radar--Satellite Fusion for Severe Convective Nowcasting with Matched Evaluation of Satellite Contributions}

\author{Dandan~Chen, Yaqiang~Wang, Anyuan~Xiong, and Enda~Zhu%
\thanks{This work has been submitted to the IEEE for possible publication. Copyright may be transferred without notice, after which this version may no longer be accessible.}%
\thanks{This research is supported by the National Natural Science Foundation of China (Grants 42450105 and 41905035), and the Science and Technology Development Foundation of Chinese Academy of Meteorological Sciences (Grant 2024KJ007).}%
\thanks{Yaqiang Wang is the corresponding author(email: yqwang@cma.gov.cn).}%
\thanks{Dandan Chen, Yaqiang Wang, and Enda Zhu are with the State Key Laboratory of Severe Weather Meteorological Science and Technology, Chinese Academy of Meteorological Sciences, Beijing, China.}%
\thanks{Dandan Chen, Yaqiang Wang and Enda Zhu are also with the Xiong'an Institute of Meteorological Artificial Intelligence, Xiong'an, China.}%
\thanks{Anyuan Xiong is with the National Meteorological Information Center, Beijing, China.}%
}

\begin{document}
\maketitle

\begin{abstract}
Radar-based nowcasting remains challenging for intense convective echoes because extrapolation cannot represent storm growth and decay, while regression-trained deep networks tend to smooth small-scale high-reflectivity structures. We present MAG-Net, a radar–satellite nowcasting framework that combines composite radar reflectivity with three Fengyun-4A (FY-4A) Advanced Geostationary Radiation Imager (AGRI) infrared fields through separate encoders and a Dual-Head decoder with Gradient-Preserving Fusion (GPF). We evaluate 73,928 spatial forecast windows, each covering a 90-minute forecast, from hourly initialization times across 122 severe-convective outbreaks over southeastern China in 2023. Matched variants separate the effects of output design and satellite input. For radar-only input, the critical success index (CSI) at 40 dBZ increases from 0.092 with regression output to 0.252 with Dual-Head/GPF. Adding AGRI fields under the same output design raises CSI at 40 dBZ to 0.273. The paired gain is 0.021, with a 95\% confidence interval of 0.018–0.024 based on outbreak-level resampling. Object-based lifecycle evaluation shows satellite gains across growth, mature, and decay stages. At the 90-minute lead, the largest relative gain occurs before the birth of a new radar core, when radar-only skill is lowest. These findings show that the output formulation and the satellite observations contribute jointly, with the larger share of the severe-echo improvement coming from the output formulation, and identify when the satellite contribution is most pronounced.
\end{abstract}


\section{Introduction}

\IEEEPARstart{S}{evere} convective precipitation events, characterized by rapid development and high intensity, pose significant threats to urban safety, aviation, and agriculture~\cite{wilson1998nowcasting,leinonen2022nowcasting,leinonen2023thunderstorm}. Precipitation nowcasting, typically defined as high-resolution forecasting with lead times of 0--2 h, is critical for mitigating these risks. However, the nonlinear growth and decay of convective cells remain a formidable challenge for operational systems~\cite{bowler2004development,bowler2006steps,pulkkinen2019pysteps}.

Operational nowcasting has traditionally relied on Numerical Weather Prediction (NWP) and radar-based extrapolation. While NWP models incorporate full atmospheric physics, they often suffer from spin-up issues and high computational latency, limiting their effectiveness for immediate localized warnings~\cite{short2022reducing,das2024hybrid}. Conversely, radar extrapolation methods based on optical flow (e.g., STEPS~\cite{bowler2006steps}) are effective for very short lead times but remain challenged in capturing convective initiation (CI) and dissipation, as they rely primarily on the persistence and advection of observed radar echoes~\cite{bowler2004development,pulkkinen2019pysteps}.

In recent years, deep learning (DL) has emerged as a powerful alternative, treating nowcasting as a spatiotemporal sequence prediction problem~\cite{de2025rainfall}. Early approaches, such as ConvLSTM~\cite{shi2015convolutional} and PredRNN~\cite{wang2017predrnn,wang2018predrnn++}, utilized recurrent units to model temporal dynamics. More recently, video-prediction and Transformer-based models have achieved strong performance in precipitation nowcasting~\cite{gao2022simvp,gao2022earthformer,zhao2024advancing}. However, regression-dominated deterministic models tend to produce blurry forecasts because pixel-wise losses favor the conditional mean and attenuate intense, small-scale echoes~\cite{chen2020deep,blau2018perception}. Physics-conditioned models such as NowcastNet~\cite{zhang2023skilful} and conditional-flow-matching models such as FlowCast~\cite{ribeiro2026flowcast} provide alternative ways to preserve sharper structures or represent forecast uncertainty. Most of these systems nevertheless remain radar-only and infer future precipitation primarily from past reflectivity.

Geostationary satellite observations provide information complementary to radar. Infrared and water-vapor measurements can characterize cloud-top evolution and the surrounding moisture field, including changes that may precede the appearance of strong radar echoes~\cite{roberts2003nowcasting,mecikalski2006forecasting}. Recognizing this potential, a growing body of research in the remote sensing community has explored multi-modal fusion for nowcasting~\cite{jin2023spatiotemporal,zheng2024cross,tan2024deep}. Recent nowcasting studies have combined radar with satellite observations~\cite{zheng2024cross,niu2024fsrgan}, or with ground-based observations~\cite{liu2025mmf,liu2025deep}. A recent example is Stormscope, a generative system that models the evolution of multispectral GOES imagery and MRMS radar reflectivity over the continental United States, with reported forecasts extending to 6~h~\cite{pathak2026learning}. Its sensors, study region, and evaluation protocol differ from the 90-min Fengyun-4A (FY-4A) task on the Convective Initiation Dataset (CIDS) considered here, so its reported scores cannot be directly ranked against ours. More broadly, existing multimodal studies vary in input products, model architectures, training objectives, spatial and temporal resolutions, and forecast protocols. These differences also make it difficult to separate improvements associated with additional observations from those associated with architecture or output design. Furthermore, attribution maps and selected cases~\cite{mamalakis2020explainable,meng2025physics,karniadakis2021physics,kashinath2021physics,li2024better} do not by themselves establish when an additional modality provides useful predictive information.

\begin{figure}[!t]
  \centering
  \includegraphics[width=\linewidth]{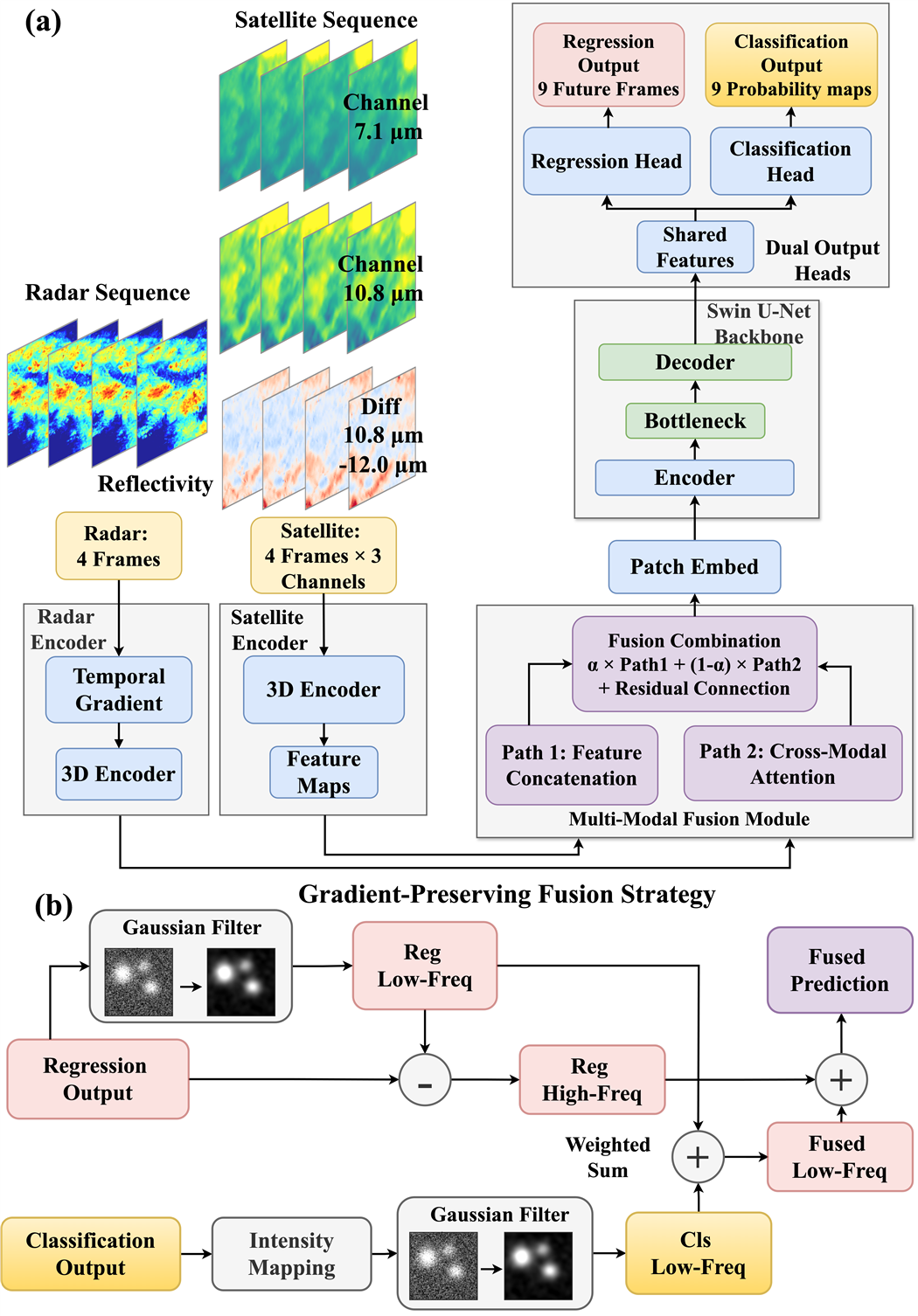}
  \caption{Overview of MAG-Net. (a) Radar and FY-4A AGRI sequences are encoded by separate streams, fused at the bottleneck, and decoded by a Swin-Transformer U-Net with a symmetric Dual-Head. The regression head predicts continuous reflectivity, whereas the classification head predicts exceedance probabilities at 12, 20, 30, and 40~dBZ. The two tasks are balanced by learned uncertainty weights. (b) Gradient-Preserving Fusion (GPF) at inference. Low-frequency components from the regression and classification-derived reflectivity fields are combined, after which the high-frequency residual of the regression output is restored. The satellite inputs are channel 10 (CH10), channel 12 (CH12), and the brightness-temperature difference BTD$_{12-13}$ (CH12 $-$ CH13). Wavelength labels on the diagram are the central wavelengths of the nominal bands defined in Section~\ref{sec:agri}.}
  \label{fig:arch}
\end{figure}

To address these issues, we develop MAG-Net (Multi-modal Attention-Guided Network, Fig.~\ref{fig:arch}) for 90-min radar--satellite nowcasting over southeastern China. MAG-Net combines radar reflectivity with three FY-4A AGRI infrared fields selected for their documented radiative sensitivities~\cite{inoue1987cloud,yang2017introducing}. Building upon the Swin-Transformer U-Net backbone of CPrecNet~\cite{park2025cprecnet}, MAG-Net retains a Dual-Stream Encoder and introduces a Symmetric Dual-Head Decoder with Gradient-Preserving Fusion (GPF). This design combines continuous reflectivity prediction with threshold-based structural constraints.

The evaluation places particular emphasis on separating the effect of the output formulation from the contribution of satellite information. Matched radar-only and multimodal variants are evaluated under consistent training and output settings, and the satellite increment is further examined across forecast lead time, convective lifecycle, and observationally defined environmental conditions. The main contributions are:

\begin{itemize}

  \item \textbf{Radar--satellite multi-modal forecasting.} MAG-Net combines radar and FY-4A AGRI observations through modality-specific encoding and explicit spatial and temporal co-registration.

  \item \textbf{Dual-Head decoding with Gradient-Preserving Fusion.} A symmetric regression/classification decoder is used to retain threshold-defined severe-echo structure while producing a continuous reflectivity forecast.

  \item \textbf{Matched separation of forecast-skill gains.} Independently trained radar-only, output-head, and satellite-channel variants are used to distinguish improvements associated with the output design from those associated with satellite observations.

  \item \textbf{Observation-conditioned evaluation.} Object-based lifecycle stratification, official-product-conditioned analyses, and Integrated Gradients~\cite{sundararajan2017axiomatic} are used to examine when the satellite increment occurs while keeping attribution-based interpretation separate from physical validation.

\end{itemize}

\section{Data and Methodology}

\subsection{Study Region and Dataset}
\label{sec:data}

The study region covers southeastern China ($104^{\circ}$E--$125^{\circ}$E, $20^{\circ}$N--$40^{\circ}$N). Radar and satellite observations are taken from CIDS~\cite{liu2026cids}. In CIDS, a regional severe convective event (SCE-R) is formed by merging temporally overlapping station-level episodes of short-duration heavy rain, thunderstorm wind, or hail into a continuous outbreak. The dataset spans from 2018 to 2023, with events from 2018--2022 used for training and 2023 events forming the evaluation cohort (Section~\ref{sec:setup}). Table~\ref{tab:data_fields} summarizes the model inputs and the products used only for post-hoc analyses.

\begin{table*}[!t]
  \centering
  \caption{OBSERVATIONS, MODEL INPUTS, AND POST-HOC ANALYSIS PRODUCTS USED IN THIS STUDY}
  \label{tab:data_fields}
  \renewcommand{\arraystretch}{1.12}
  \small

  \textit{(a) Fields used by the forecasting models}\par\smallskip
  \begin{tabularx}{\textwidth}{@{}>{\raggedright\arraybackslash}p{0.11\textwidth}>{\raggedright\arraybackslash}p{0.2\textwidth}l>{\raggedright\arraybackslash}X>{\raggedright\arraybackslash}X@{}}
    \toprule
    Field & Spectral band / product & Native grid & Observational role & Use in this study \\
    \midrule
    Radar CR &
      CMA composite reflectivity &
      $2001\times2101$, $0.01^{\circ}$ &
      Precipitation-echo intensity and motion &
      Input and prediction target \\
    AGRI CH10 &
      6.9--7.3~$\mu$m (mid-level water vapor) &
      $501\times526$, $0.04^{\circ}$ &
      Mid-tropospheric moisture context &
      Default MAG-Net; five-band variant \\
    AGRI CH12 &
      10.3--11.3~$\mu$m (infrared window) &
      $501\times526$, $0.04^{\circ}$ &
      Cloud-top thermal structure and cooling &
      Default MAG-Net; five-band variant \\
    BTD$_{12-13}$ &
      CH12 $-$ CH13 (11.5--12.5~$\mu$m) &
      $501\times526$, $0.04^{\circ}$ &
      Split-window cloud-radiative contrast &
      Default MAG-Net; five-band variant \\
    Additional AGRI fields &
      CH09 (5.8--6.7~$\mu$m), CH10, CH12, CH13, CH14 (13.2--13.8~$\mu$m); BTD$_{12-13}$, BTD$_{09-12}$, BTD$_{10-12}$, BTD$_{12-14}$ &
      $501\times526$, $0.04^{\circ}$ &
      Broader cloud and water-vapor context &
      Five-band ablation only \\
    \bottomrule
  \end{tabularx}

  \vspace{0.55em}
  \textit{(b) Co-registration and products used only in post-hoc analyses}\par\smallskip
  \begin{tabularx}{\textwidth}{@{}>{\raggedright\arraybackslash}p{0.14\textwidth}>{\raggedright\arraybackslash}X>{\raggedright\arraybackslash}X l@{}}
    \toprule
    Item & Definition / quality control & Role in analysis & Network input \\
    \midrule
    Radar--AGRI pairing &
      Common domain; paired native $64\times64$ AGRI and $256\times256$ radar crops; bilinear resampling &
      Spatial and temporal co-registration &
      Preprocessing \\
    Official CIX &
      Latest convective-initiation product (CIX) flag at or before initialization, confirmed by observed $\geq$30-dBZ radar echo within 20~km &
      CIX-conditioned consistency analysis &
      No \\
    FY-4B GIIRS AII K-index &
      Geostationary Interferometric Infrared Sounder (GIIRS) Atmosphere Instability Index (AII); data quality flag (DQF)$\leq1$; summarized by outbreak &
      Environmental-instability stratification &
      No \\
    FY-4 LMI &
      Lightning Mapping Imager (LMI) seasonal lightning-density field &
      Spatial comparison with severe-echo occurrence &
      No \\
    \bottomrule
  \end{tabularx}
\end{table*}

\subsection{Radar Reflectivity}

We utilize CIDS composite-reflectivity (CR) mosaics from the China Meteorological Administration (CMA) operational network~\cite{bai2020image,liu2026cids}. The data are provided on a regular $0.01^{\circ}$ grid ($2001\times2101$) at 10-min intervals, with invalid pixels masked. Reflectivity $Z$ (dBZ) is mapped to $x_{rad}\in[0,1]$ by

\begin{equation}
x_{rad} =
\frac{\mathrm{clip}(Z, Z_{min}, Z_{max}) - Z_{min}}
{Z_{max} - Z_{min}},
\end{equation}

with $Z_{min}=10$~dBZ and $Z_{max}=50$~dBZ. The upper bound reduces the influence of rare extreme values during training. All reported verification scores are computed in dBZ after inverting this transformation.

\subsection{FY-4A AGRI Observations and Channel Roles}
\label{sec:agri}

FY-4A is a Chinese geostationary meteorological satellite. Its Advanced Geostationary Radiation Imager (AGRI) observes the Earth in 14 visible, near-infrared, and infrared bands~\cite{yang2017introducing}. The water-vapor and thermal-infrared bands used here have a nominal spatial resolution of approximately 4~km. We use the AGRI brightness-temperature (BT) fields distributed with CIDS on a $0.04^{\circ}$ grid ($501\times526$), covering the same geographic domain as the radar mosaics. The wavelengths, native grids, and roles of the selected fields are summarized in Table~\ref{tab:data_fields}. We select three infrared fields for the default MAG-Net:

\begin{itemize}

  \item \textbf{CH10} (6.9--7.3~$\mu$m; central wavelength 7.1~$\mu$m), a mid-level water-vapor band that provides information on the mid-tropospheric moisture environment.

  \item \textbf{CH12} (10.3--11.3~$\mu$m; central wavelength 10.8~$\mu$m), an infrared-window band that provides cloud-top brightness temperature. Temporal cooling can accompany convective cloud development~\cite{roberts2003nowcasting,mecikalski2006forecasting}.

  \item \textbf{BTD$_{12-13}$} (CH12 $-$ CH13), where CH13 is the 11.5--12.5~$\mu$m split-window band (central wavelength 12.0~$\mu$m). This brightness-temperature difference is sensitive to optical thickness and semi-transparent cirrus, but that sensitivity is not specific to cloud phase~\cite{inoue1987cloud,mayer2024well}.

\end{itemize}

These radiance-derived fields are used as observational indicators rather than as direct retrievals of atmospheric stability or convective initiation. The three fields are clipped to $[200,280]$~K, $[200,320]$~K, and $[-10,10]$~K, respectively, and linearly scaled to $[0,1]$. Only thermal-emission channels are used in the default configuration so that the inputs retain a consistent physical interpretation between day and night. Shortwave bands near 3.7~$\mu$m contain both reflected solar and emitted thermal contributions.

A five-band variant uses CH09 (5.8--6.7~$\mu$m, upper-level water vapor), CH10, CH12, CH13, and CH14 (13.2--13.8~$\mu$m), together with four brightness-temperature differences: BTD$_{12-13}$, BTD$_{09-12}$ (CH09 $-$ CH12), BTD$_{10-12}$ (CH10 $-$ CH12), and BTD$_{12-14}$ (CH12 $-$ CH14). These nine fields are trained separately to test whether a broader radiance set improves forecast skill (Section~\ref{sec:baselines}). Official stability and convective-initiation products are not used as model inputs. The AGRI cloud optical thickness and effective radius are produced by a separate bi-spectral reflectance retrieval~\cite{liu2023cloud} and are likewise excluded.

\subsection{Radar--AGRI Co-registration}
\label{sec:coreg}

\begin{figure}[!t]
  \centering
  \includegraphics[width=\linewidth]{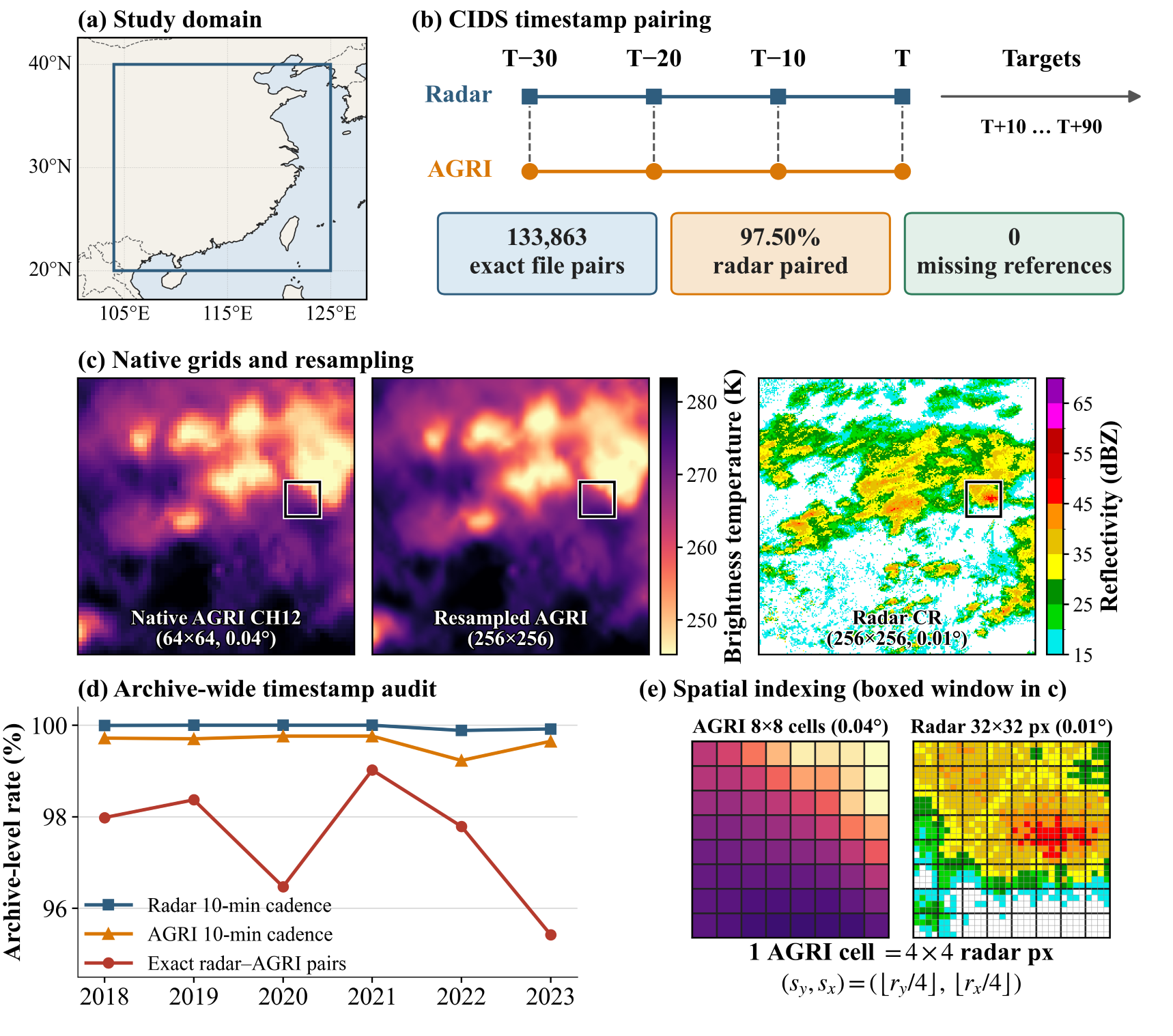}
  \caption{Radar--AGRI co-registration and pairing audit. (a) Study domain and an example $256\times256$ radar crop. (b) Temporal pairing of four radar--AGRI input frames and nine radar forecast targets, with archive-wide pairing statistics. (c) Native $64\times64$ AGRI CH12 crop ($0.04^{\circ}$), its bilinearly resampled $256\times256$ field, and the corresponding radar CR crop ($0.01^{\circ}$). (d) Yearly availability of radar and AGRI frames and the fraction of radar frames with an exact AGRI timestamp match. (e) Spatial indexing of the boxed region in (c), where one native AGRI cell corresponds to $4\times4$ radar pixels. Resampling aligns the input grids but does not increase the effective spatial resolution of AGRI.}
  \label{fig:coreg}
\end{figure}

Radar and AGRI observations are aligned explicitly in both space and time before entering the network (Fig.~\ref{fig:coreg}). The two products share the same geographic domain and are distributed on north-up grids with a nominal resolution ratio of 4:1. Accordingly, one native AGRI grid cell corresponds to $4\times4$ radar pixels, and the full arrays share their endpoint coordinates. Training and cohort-evaluation windows use a fixed stride of 192 radar pixels starting at index 0, so every radar origin $(r_y,r_x)$ is divisible by 4. The corresponding native AGRI crop is $64\times64$ and begins at $(r_y/4,r_x/4)$; the remainders $r_y\bmod 4$ and $r_x\bmod 4$ are zero for every such window. The AGRI crop is then bilinearly resampled to $256\times256$ by mapping the first and last pixels of the $64\times64$ crop onto the first and last pixels of the radar window. Because a 64-pixel AGRI crop spans 252 radar-pixel steps while the radar window spans 255, this endpoint mapping shifts the far edge by 3 radar pixels relative to placing each AGRI sample on every fourth radar pixel. The resampling matches the tensor dimensions but does not increase the information content of the satellite observations, whose effective spatial footprint remains approximately 4~km.

We use the CIDS-aligned timestamps as distributed and apply no temporal interpolation~\cite{liu2026cids}. Sensitivity to residual spatial and temporal misalignment is evaluated separately in Section~\ref{sec:res_coreg}.

\subsection{Problem Formulation}

Precipitation nowcasting is formulated as a spatiotemporal sequence prediction problem~\cite{shi2017deep}. Given historical radar sequences
$\mathcal{X}_{rad}\in\mathbb{R}^{T_{in}\times H\times W}$
and satellite sequences
$\mathcal{X}_{sat}\in\mathbb{R}^{T_{in}\times C_{sat}\times H\times W}$,
the goal is to predict the future radar reflectivity sequence
$\hat{\mathcal{Y}}\in\mathbb{R}^{T_{out}\times H\times W}$.
In this study, $T_{in}=4$, $T_{out}=9$, $H=W=256$, and $C_{sat}=3$ for the default configuration, corresponding to inputs at $T-30$, $T-20$, $T-10$, and $T$ min and forecasts from $T+10$ to $T+90$ min at 10-min intervals.

\subsection{MAG-Net Architecture}

\label{sec:architecture}

As illustrated in Fig.~\ref{fig:arch}, MAG-Net adopts a Swin-Transformer U-Net backbone, extending the architecture of CPrecNet~\cite{park2025cprecnet} to a multi-modal context. The network consists of a Dual-Stream Encoder, a Cross-Modal Fusion Module, and a Symmetric Dual-Head Decoder.

\subsubsection{Dual-Stream Hierarchical Encoder}

To handle the heterogeneous statistical properties of radar and satellite data, we design two parallel encoding streams. Dual-stream designs have been used previously for spatiotemporal forecasting~\cite{chen2026synqpf}. The \textbf{Radar Stream} captures short-term evolution by computing temporal gradients via frame differencing
($\Delta\mathcal{X}_{t}=\mathcal{X}_{t}-\mathcal{X}_{t-1}$)
and stacking them with the raw intensity frames. A 3D convolutional block extracts spatiotemporal features, which are then spatially downsampled to a compact latent resolution ($32\times32$).

The \textbf{Satellite Stream} processes the aligned four-frame history of the selected AGRI fields using a dedicated 3D convolutional encoder and projects the resulting features to the same latent space as the radar features.

\subsubsection{Cross-Modal Attention Fusion}

Deep fusion is performed at the bottleneck level. We employ a Cross-Modal Attention mechanism where radar features $\mathbf{F}_{rad}$ serve as the Query, while satellite features $\mathbf{F}_{sat}$ serve as both Key and Value. To optimize memory efficiency, the fusion operates at reduced resolution. The attention branch is combined with a convolutional concatenation path:

\begin{equation}
\begin{aligned}
\mathbf{F}_{fused} =\;& \mathbf{F}_{rad}
+ \alpha \cdot \mathrm{Attention}(\mathbf{Q}_{rad}, \mathbf{K}_{sat}, \mathbf{V}_{sat}) \\
&+ (1-\alpha) \cdot [\mathbf{F}_{rad}; \mathbf{F}_{sat}]_{conv},
\end{aligned}
\end{equation}

where $[\cdot;\cdot]_{conv}$ denotes concatenation followed by a $1\times1$ convolution, and $\alpha$ is a learnable scalar initialized to 0.5. Since fusion operates at the reduced latent resolution, the fused representation is subsequently upsampled to the input resolution through three transposed-convolution layers before being passed to the Swin-Transformer U-Net~\cite{liu2021swin}.

\subsection{Multi-Task Learning Strategy}

\label{sec:loss}

To address the excessive smoothing of regression-based nowcasting, MAG-Net employs a symmetric Dual-Head decoder. The \textbf{Regression Head} outputs continuous reflectivity values $\hat{\mathcal{Y}}_{reg}$, while the \textbf{Classification Head} predicts probability maps for four ordered intensity thresholds ($12$, $20$, $30$, and $40$~dBZ). The auxiliary classification task provides explicit supervision on threshold-defined echo structures, complementing the continuous regression objective.

We employ homoscedastic uncertainty weighting to dynamically balance the two tasks~\cite{kendall2018multi}. The total loss $\mathcal{L}_{total}$ is defined as follows:

\begin{equation}
\label{eq:multitask_loss}
\mathcal{L}_{total}
=
\exp(-s_1)\mathcal{L}_{reg}
+
\exp(-s_2)\mathcal{L}_{cls}
+
s_1+s_2,
\end{equation}

where $s_i=\log\sigma_i^{2}$ are learnable log-variance parameters. $\mathcal{L}_{reg}$ uses Balanced MSE (BMSE)~\cite{ren2022balanced} to increase the contribution of high-reflectivity pixels. $\mathcal{L}_{cls}$ combines Dice loss~\cite{milletari2016v} and Binary Cross-Entropy (BCE), computed separately for each lead time and threshold before aggregation. Lead-time weights increase linearly from 1 at 10~min to 2 at 90~min. The 40-dBZ threshold receives a class weight of 2, and positive pixels at 30 and 40~dBZ receive BCE weights of 2 and 4, respectively, to reduce class imbalance.

\subsection{Inference-Time Gradient-Preserving Fusion (GPF)}

\label{sec:gpf}

Standard regression models tend to produce overly smooth forecasts, whereas the classification head retains stronger threshold-defined structure but does not directly produce a continuous reflectivity field. To combine these complementary outputs, we use the Gradient-Preserving Fusion (GPF) strategy shown in Fig.~\ref{fig:arch}(b).

First, the classification probabilities are mapped to a pseudo-reflectivity field without introducing additional learned parameters. Let
$p_c=\mathrm{sigmoid}(\ell_c)$
denote the probability that reflectivity exceeds threshold $t_c$. We treat $\{p_c\}$ as discrete samples of the survival function
$S(t)=P(Y\geq t)$,
with $S(Z_{min})=1$, and approximate the normalized expected reflectivity by

\begin{equation}
\hat{Y}_{map} = \sum_{k=0}^{K} w_k\,\bar{S}_k,
\end{equation}

where the intervals are bounded by $Z_{min}$, the four thresholds, and $Z_{max}$. Here $w_k$ is the normalized width of interval $k$ and $\bar{S}_k$ is the mean survival probability at its two endpoints under the trapezoidal approximation.

Next, we apply a Gaussian low-pass filter $G_\sigma$ to the regression output $\hat{Y}_{reg}$ and the classification-derived field $\hat{Y}_{map}$:

\begin{equation}
Y_{low}^{reg}=G_\sigma(\hat{Y}_{reg}), \qquad
Y_{low}^{cls}=G_\sigma(\hat{Y}_{map}).
\end{equation}

The high-frequency component is isolated from the regression output as

\begin{equation}
Y_{high}^{reg}=\hat{Y}_{reg}-Y_{low}^{reg}.
\end{equation}

Finally, the fused prediction is obtained by combining the low-frequency structure with the preserved regression details:

\begin{equation}
\hat{Y}_{fused}
=
(1-\lambda)Y_{low}^{cls}
+
\lambda Y_{low}^{reg}
+
Y_{high}^{reg},
\end{equation}

where $\lambda=0.5$ controls the structural mixing weight and $\sigma=3$ pixels determines the Gaussian filtering scale. GPF operates only at inference time and introduces no additional learned parameters. Its effect relative to the individual regression and classification outputs is evaluated in Section~\ref{sec:res_ablation}.

\section{Experiments}

\label{sec:experiments}

\subsection{Training and Checkpoint Selection}

\label{sec:setup}

All learning-based models are implemented in PyTorch. MAG-Net and its matched variants are trained on the 2018--2022 SCE-R events. We use the Adam optimizer with a learning rate of $5\times10^{-4}$, a batch size of 16, and FP32 training on a single NVIDIA H100 GPU for 40 epochs.

Model selection is based on a validation score defined as BMSE plus 0.1 times the Dice loss.

For the uncertainty-weighted multi-task objective, the task log-variances are initialized at $s_i=0.5$ and constrained to $[-1,2]$ to prevent either task from being effectively suppressed. Across the Dual-Head models, the learned weights remain comparable between the regression and classification objectives.

All matched MAG-Net variants follow the same selection protocol. Single-head variants are selected using the validation loss of their respective output head. The controlled comparisons among these variants therefore use a consistent model-selection procedure.

\subsection{Evaluation Cohort and Metrics}

\label{sec:metrics}

For evaluation, the 2023 manifests retain one initialization per hour, and each initialization contributes multiple $256\times256$ spatial windows. Each sequence contains four input frames and nine forecast targets at 10-min intervals. The cohort contains 122 regional severe-convective outbreaks across spring (MAM, 63), summer (JJA, 35), and autumn (SON, 24), with no winter outbreaks, yielding 73,928 evaluated windows under the same spatial-window protocol for all models.

Forecast skill is evaluated using the critical success index (CSI), probability of detection (POD), and false alarm ratio (FAR) at 12, 20, 30, and 40~dBZ. We also report the fractions skill score (FSS)~\cite{roberts2008scale} with a 5-km neighborhood, root-mean-square error (RMSE), and mean absolute error (MAE) in dBZ. The 12-dBZ threshold approximately separates precipitation from non-precipitation near the lower normalization bound.

Categorical metrics are computed from contingency counts accumulated across the evaluation cohort separately at each forecast lead. Tabulated values report the mean over the nine leads. Because the study focuses on the retention of intense convective echoes, CSI$_{40}$ and FSS at 40~dBZ are used as the primary severe-echo metrics. RMSE and MAE are reported alongside them to quantify the corresponding behavior in pixel-wise error.

\subsection{Statistical Testing}

\label{sec:stats}

Spatial windows drawn from the same convective outbreak are not statistically independent. We therefore quantify uncertainty using a paired cluster bootstrap~\cite{efron1993introduction}, with the outbreak treated as the resampling unit for the main model comparisons.

For each of 10,000 bootstrap resamples, the 122 outbreaks are sampled with replacement, and the contingency counts and continuous-error statistics of the paired models are re-accumulated over exactly the same resampled outbreaks. Performance metrics and their paired differences are then recomputed. We report percentile 95\% confidence intervals and regard a paired difference as statistically supported when its 95\% confidence interval excludes zero.

The same resampling principle is used for the subsequent analyses. Tracked objects are used as the resampling unit for the lifecycle evaluation (Section~\ref{sec:lifecycle}), whereas outbreaks are used for the product-conditioned, attribution, and seasonal analyses. The corresponding resampling unit is stated in each figure caption.

\subsection{Baselines and Controlled Variants}

\label{sec:baselines}

The experimental design serves two distinct purposes. External radar-only baselines evaluate the overall forecast performance of MAG-Net relative to representative nowcasting paradigms, whereas matched MAG-Net variants are used to separate improvements associated with the output formulation from those associated with satellite information.

\subsubsection{External baselines}

We compare MAG-Net with four radar-only models representing extrapolation, deterministic deep learning, Transformer-based forecasting, and flow matching:

\begin{itemize}

  \item \textbf{pySTEPS}~\cite{pulkkinen2019pysteps}: Lucas--Kanade optical flow with semi-Lagrangian advection and no trainable parameters.

  \item \textbf{SimVP-v2}~\cite{gao2022simvpv2}: a convolutional video-prediction model based on gated spatiotemporal attention (gSTA), implemented with OpenSTL~\cite{tan2023openstl}.

  \item \textbf{EarthFormer}~\cite{gao2022earthformer}: a cuboid-attention Transformer developed for Earth-system forecasting.

  \item \textbf{FlowCast}~\cite{ribeiro2026flowcast}: a conditional-flow-matching model with its matched autoencoder. Deterministic verification uses the mean of eight generated members.

\end{itemize}

All learning-based baselines are retrained on the 2018--2022 events using the same four-input, nine-output, 10-min forecast protocol as MAG-Net.

\subsubsection{Controlled variants}

Matched variants are used to isolate two components of the proposed system: output formulation and satellite input.

\begin{itemize}

  \item \textbf{Output formulation.} The radar-only Dual-Head model (RD-Dual) removes the satellite stream while retaining the Dual-Head/GPF output design. Radar-only Regression uses the corresponding regression output and matches the CPrecNet-style architecture and prediction target. Multimodal Regression and Multimodal Classification are independently trained single-head models. MAG-Net (regression head) is the regression output of the Dual-Head MAG-Net before GPF, evaluated from the same checkpoint rather than a separately trained model.

  \item \textbf{Satellite input.} Radar+IR uses the infrared-window band CH12 alone. Radar+IR+WV adds the mid-level water-vapor band CH10. The default MAG-Net uses CH10, CH12, and BTD$_{12-13}$, while the five-band AGRI variant uses the nine fields defined in Section~\ref{sec:agri}. Each configuration is independently retrained, rather than evaluated through inference-time channel masking.

\end{itemize}

Together, these variants allow the performance difference between the full MAG-Net and radar-only forecasting to be decomposed into contributions from the output design and the additional satellite observations.

\subsection{Object-Based Lifecycle Stratification}

\label{sec:lifecycle}

Aggregate forecast scores cannot determine whether the satellite increment depends on the stage of convective evolution. We therefore perform an object-based lifecycle analysis in which the 2023 forecasts are stratified by stages diagnosed retrospectively from observed cloud objects and their associated radar cores (Figs.~\ref{fig:tracking} and~\ref{fig:stages}). Object-based mesoscale convective system (MCS) analyses have also been used to relate tracked convective systems to regional heavy rainfall~\cite{jiang2026tracking}. The lifecycle labels are used only for evaluation and are never provided to either forecasting model.

\subsubsection{Tracking and matching}

Satellite infrared object tracking has been used to construct long-term MCS datasets~\cite{huang2018long}. Cold-cloud objects are identified from AGRI CH12 as contiguous regions with brightness temperature below 240~K and area exceeding 200~km$^2$. These objects are tracked using the algorithm of Chen et al.~\cite{chen2019mesoscale}, with its original parameters retained. Consecutive objects are associated primarily by spatial overlap. For small unmatched objects, a single Kalman-filter position forecast is used as a fallback. A candidate pair is accepted only when the implied speed does not exceed 5~km~min$^{-1}$, the area ratio does not exceed 4, and the minimum-BT difference does not exceed 20~K. Objects involved in split or merge events are excluded from this fallback association.

\begin{figure}[!t]

  \centering

  \includegraphics[width=\linewidth]{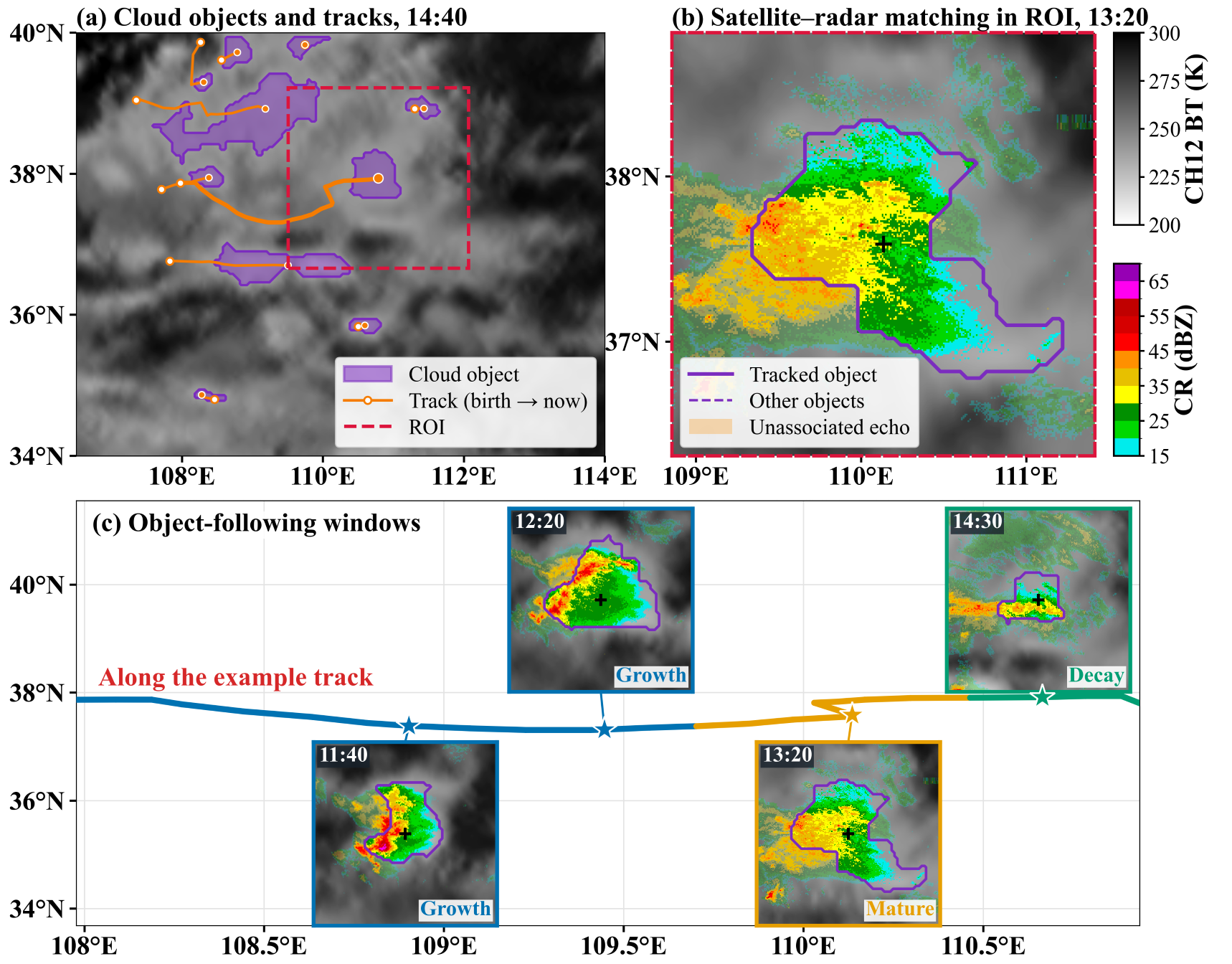}

  \caption{Cloud-object tracking and satellite--radar matching used for lifecycle evaluation. (a) CH12 cold-cloud objects ($<240$~K and $>200$~km$^2$) and their tracks from object birth to the current time. (b) Radar reflectivity associated with the tracked cloud object within the region of interest. Colored echoes are included in the object association, whereas shaded echoes are excluded. (c) Object-following $256\times256$ evaluation windows at four issue times along the example track, labeled by the retrospectively diagnosed lifecycle stage.}

  \label{fig:tracking}

\end{figure}

Radar convective cores are then identified within each tracked cloud object as contiguous regions with reflectivity $\geq35$~dBZ and an area of at least 16 native radar pixels (Fig.~\ref{fig:tracking}(b)). This 35-dBZ level is the conventional convective-core threshold used for the lifecycle and convective-initiation definitions below. The classification head predicts exceedance at 12, 20, 30, and 40~dBZ, so scores at 35~dBZ are computed from the continuous reflectivity forecast. For each forecast initialization along a track, a $256\times256$ radar window is centered on the tracked object and kept fixed over the full 90-min forecast period (Fig.~\ref{fig:tracking}(c)). These windows are not restricted to the stride-192 grid. The paired AGRI crop uses the native pixel nearest the same center, which can leave an origin offset of up to 2 radar pixels relative to an exact factor-of-4 index, and is bilinearly resized to $256\times256$. MAG-Net and the matched RD-Dual model are evaluated within exactly the same windows, so their difference is not affected by changes in the evaluation domain.

\subsubsection{Stage definition}

Lifecycle stages are defined from the observed evolution of the 35-dBZ radar-core area, following the growth--mature--decay decomposition used in MCS studies~\cite{feng2018structure,feng2021global}. For each track, the core area $A(t)$ is normalized by its track maximum $A_{max}$, and time is normalized to the track duration $\tau\in[0,1]$. A penalized cubic smoothing spline is fitted to $A/A_{max}$, with the smoothing strength selected by generalized cross-validation.

\begin{figure}[!t]

  \centering

  \includegraphics[width=\linewidth]{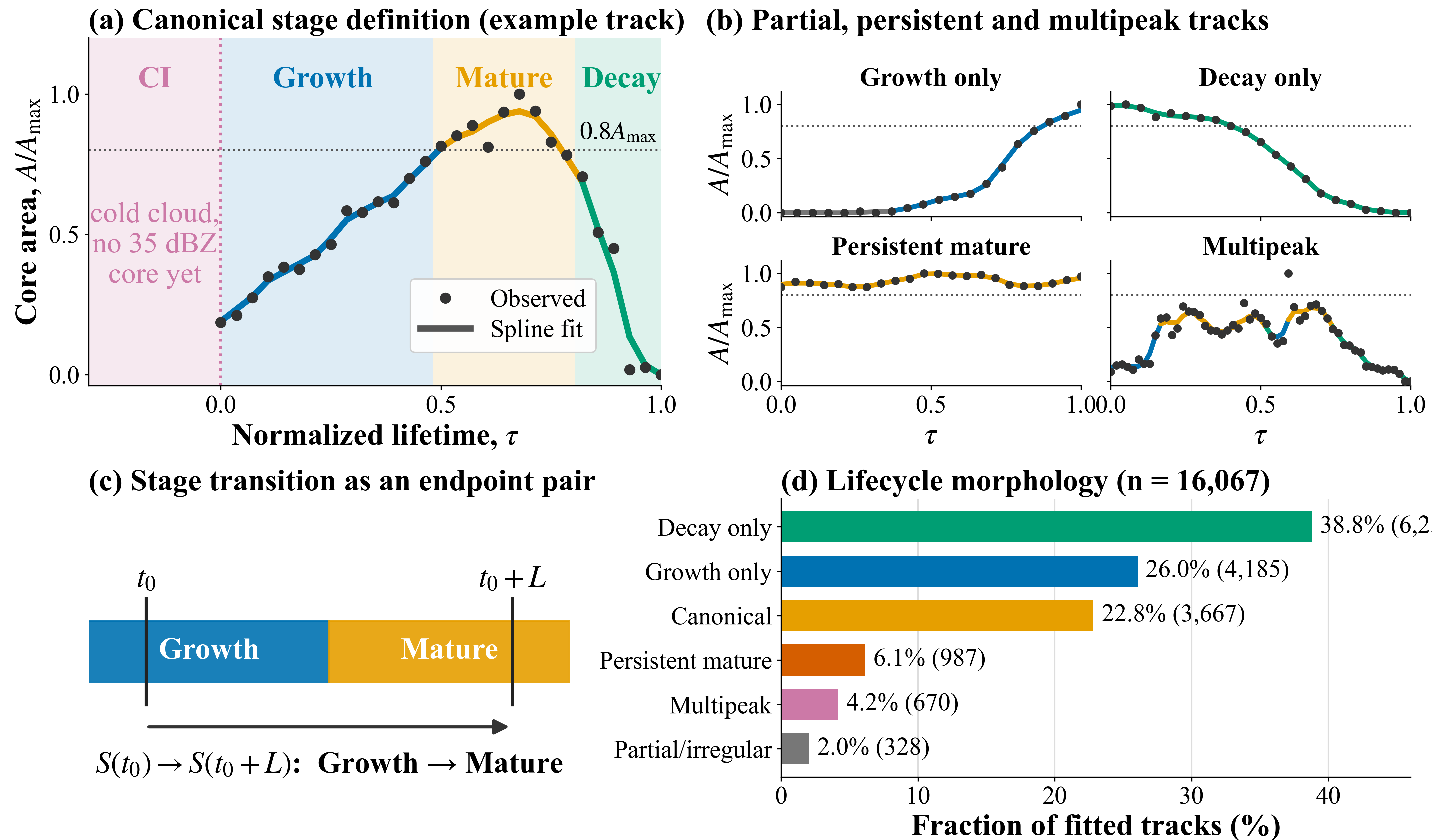}

  \caption{Definition of convective lifecycle stages. (a) Canonical track showing observed 35-dBZ core area normalized by its track maximum, the fitted smoothing spline, and Growth, Mature, and Decay stages defined by the $0.8A_{\max}$ crossings. CI denotes qualifying pre-core frames within 90~min before the first persistent 35-dBZ core. (b) Examples of noncanonical tracks, including growth-only, decay-only, persistent-mature, and multipeak evolution, with stages assigned using the corresponding rules. (c) Issue-time stage $S(t_0)$ and valid-time stage $S(t_0+L)$ used in the transition analysis. (d) Morphological composition of the 16,067 fitted tracks in the 2023 cohort. Lifecycle labels are diagnosed retrospectively and are not provided to the forecasting models.}

  \label{fig:stages}

\end{figure}

For a canonical lifecycle, Growth extends from core formation until the fitted area first reaches $0.8A_{max}$, Mature spans the interval over which the fitted area remains above this threshold, and Decay follows the final downward crossing (Fig.~\ref{fig:stages}(a)). Convective initiation (CI) is treated separately as a pre-core event within the tracked cold-cloud object~\cite{feng2021global}: it consists of zero-core frames within 90~min before the first appearance of a new 35-dBZ core~\cite{mecikalski2006forecasting} that persists for at least three frames and is not attributable to advection from outside the tracked object or to splitting from an existing core, following newness constraints used in radar CI studies~\cite{bai2020image}.

Real convective tracks do not always exhibit a complete growth--mature--decay sequence. Among the 16,067 fitted tracks, 22.8\% are canonical, while the remainder include growth-only, decay-only, persistent-mature, multipeak, and irregular trajectories (Fig.~\ref{fig:stages}(b), (d)). These noncanonical tracks are retained rather than discarded. Growth-only and decay-only tracks are assigned stages according to the sign of the fitted area tendency, persistent-mature tracks remain above $0.8A_{max}$, and multipeak tracks are divided at fitted valleys so that distinct peaks can be treated as separate lifecycle cycles. Flat or ambiguous portions are labeled as transition. Importantly, a Mature stage is not inferred solely because a truncated track reaches its local maximum.

Stage labels are assigned retrospectively from the complete observed track. For each forecast issue time $t_0$, we record both the issue-time stage $S(t_0)$ and the valid-time stage $S(t_0+L)$, allowing the analysis to distinguish stage persistence from lifecycle transitions during the forecast (Fig.~\ref{fig:stages}(c)). For example, Growth$\rightarrow$Mature denotes a forecast initialized during growth whose verifying observation is in the mature stage.

To reduce dependence among strongly overlapping 10-min forecast issues, one midpoint issue is retained from each contiguous lifecycle-stage episode. All CI episodes are retained, with one initialization selected from each 30-min pre-birth interval, while Growth, Mature, and Decay are limited to a seeded random sample of at most 3,000 episodes per stage (seed 42). The resulting evaluation set contains 8,735 issues from 6,361 tracks in 108 outbreaks. MAG-Net is compared with the matched RD-Dual model using CSI at 35~dBZ on the continuous reflectivity forecast, with uncertainty estimated from 2,000 paired bootstrap resamples of tracked objects.

\subsection{Observation-Conditioned Analyses and Attribution Diagnostics}

\label{sec:products}

The matched comparison between MAG-Net and RD-Dual quantifies the average contribution of satellite input, but does not indicate whether this increment is associated with particular convective environments or spatial patterns. We therefore relate the satellite increment to several FY-4 observational products that are not used as model inputs, and separately use Integrated Gradients (IG) as a model-sensitivity diagnostic.

\subsubsection{Observation-conditioned analyses}

Three FY-4 products are used to characterize the conditions under which the satellite increment occurs:

\begin{itemize}

  \item \textbf{Official AGRI convective-initiation product (CIX).}
  For each hourly initialization, we use the latest official CIX flag available at or before the issue time. An outbreak is retained when at least one flagged object is subsequently associated with observed radar reflectivity $\geq30$~dBZ within 20~km. This criterion retains 105 of the 122 outbreaks (73,350 sequences). Forecast scores are evaluated over the full spatial windows of the selected outbreaks rather than only at CIX locations. Because CIX is itself derived from AGRI observations, this analysis is treated as an official-product-conditioned consistency check rather than as validation from an independent sensor.

  \item \textbf{FY-4B Geostationary Interferometric Infrared Sounder (GIIRS) Atmosphere Instability Index (AII) K-index.}
  For each outbreak, we compute the median quality-controlled K-index (data quality flag, DQF$\leq1$) over the outbreak domain. The 120 outbreaks with valid retrievals are divided into tertiles using cut points of 14.1 and 24.0~K. Because the K-index is retrieved by a different instrument on a different satellite from the AGRI inputs, it provides an external description of the large-scale thermodynamic environment. We compare the increment of MAG-Net over RD-Dual across the three instability strata.

  \item \textbf{FY-4 Lightning Mapping Imager (LMI).}
  Seasonal lightning-density fields are used as an independent indicator of the spatial distribution of convective activity. These fields are compared with the observed occurrence of 40-dBZ echoes and with the spatial distribution of the forecast difference between MAG-Net and RD-Dual.

\end{itemize}

For the K-index analysis, we additionally test whether the satellite increment differs between the high- and low-instability tertiles using the outbreak-cluster bootstrap.

\subsubsection{Integrated-Gradients diagnostic}

Integrated Gradients~\cite{sundararajan2017axiomatic} is used to examine the sensitivity of the MAG-Net checkpoint to its radar and satellite inputs. The attribution target is the spatial sum of the predicted exceedance probabilities at 20, 30, or 40~dBZ for a given forecast lead.

Radar and satellite inputs are integrated separately from a zero baseline in normalized space while the other modality is held at its observed value. Attributions are approximated using a 50-step midpoint Riemann sum. We summarize the resulting sensitivities using three diagnostics: (i) the satellite share of absolute attribution, defined as the per-element mean absolute satellite attribution divided by the sum of the corresponding radar and satellite means, (ii) the distribution of satellite attribution among CH10, CH12, and BTD$_{12-13}$, and (iii) the relative completeness error,
$|\sum\mathrm{IG}-(F(x)-F(x_0))|/|F(x)-F(x_0)|$.

The analysis uses 500 randomly sampled 2023 sequences from 72 outbreaks, evaluated at all nine forecast leads and all three attribution thresholds. Confidence intervals are estimated from 2,000 outbreak-cluster bootstrap resamples.

IG is interpreted only as a sensitivity diagnostic for the selected checkpoint and baseline. It does not quantify the unique information contributed by an input variable and is not used as evidence that the model has learned a physical mechanism.

\section{Results}
\label{sec:results}

\subsection{Comparison with External Baselines}

\label{sec:res_overall}

\begin{figure}[!t]
  \centering
  \includegraphics[width=\linewidth]{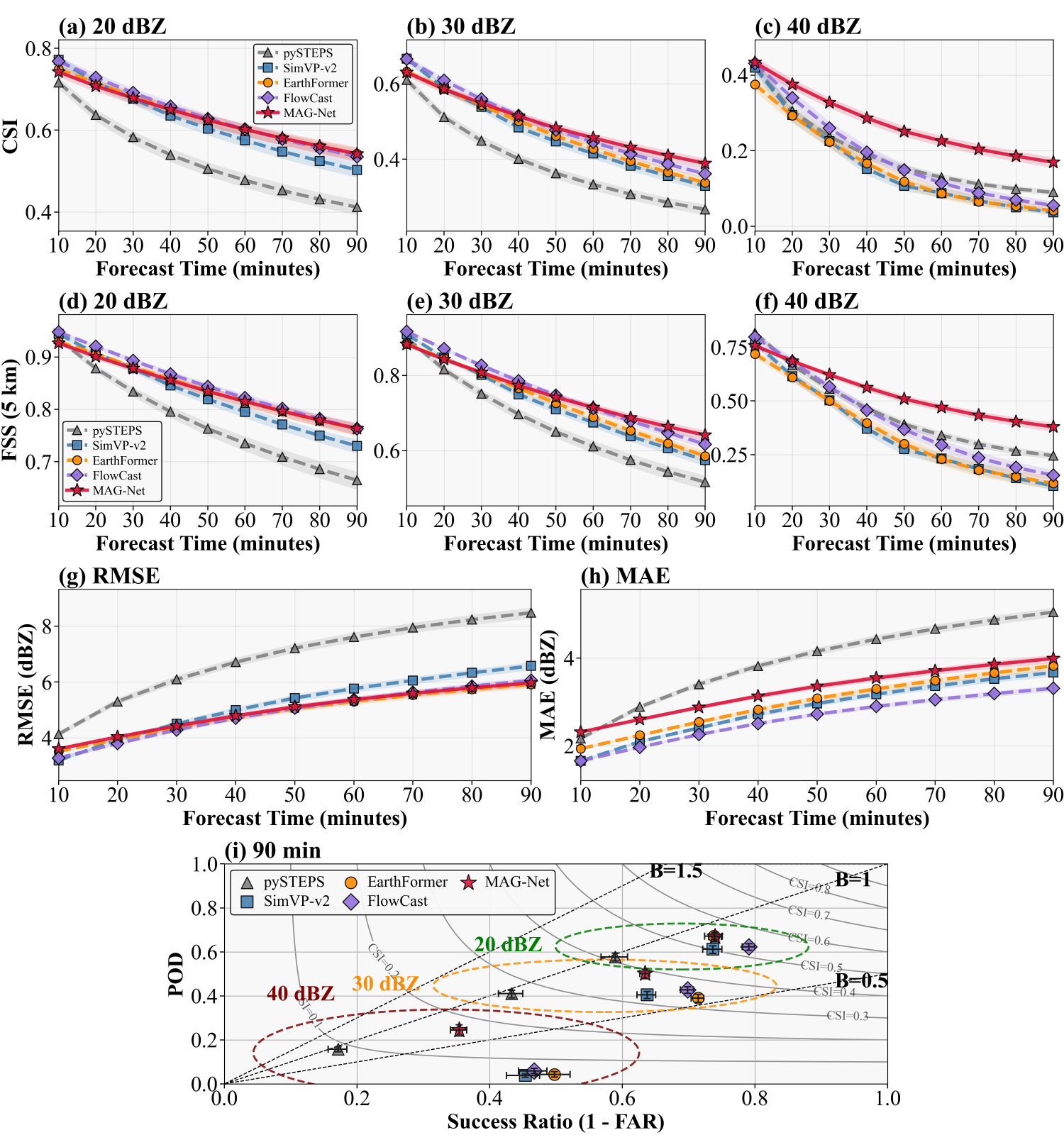}
  \caption{Lead-time performance of MAG-Net and external radar-only baselines on the 2023 evaluation cohort (73,928 spatial windows from 122 SCE-R outbreaks). (a)--(c) CSI at 20, 30, and 40~dBZ. (d)--(f) FSS at a 5-km neighborhood scale for the same thresholds. (g) RMSE and (h) MAE. (i) Performance diagram at 90~min for 20, 30, and 40~dBZ. Diagonal lines denote frequency bias, and curved contours denote CSI. Shading in (a)--(h) and error bars in (i) denote the 95\% confidence interval of each score from the outbreak-cluster bootstrap (10,000 resamples). MAG-Net uses the GPF output, and FlowCast uses the mean of eight generated members.}
  \label{fig:overall}
\end{figure}

Fig.~\ref{fig:overall} and Table~\ref{tab:performance_ablation}(a) compare MAG-Net with the external radar-only baselines. MAG-Net achieves higher CSI$_{40}$ (0.273) and 40-dBZ FSS (0.536) than all compared baselines, with the separation at 40~dBZ increasing toward longer lead times (Fig.~\ref{fig:overall}(c), (f)). Relative to FlowCast, which has the highest external CSI$_{40}$ among the compared baselines, the paired difference is $+0.084$ [0.077, 0.091] for CSI$_{40}$ and $+0.120$ [0.107, 0.132] for 40-dBZ FSS.

The same advantage is not observed across all metrics. FlowCast has the highest CSI$_{12}$ (not shown) and CSI$_{20}$ and the lowest RMSE and MAE, while EarthFormer and SimVP-v2 remain competitive at lower reflectivity thresholds. MAG-Net's CSI$_{30}$ is statistically indistinguishable from that of FlowCast ($+0.002$ [$-0.001$, 0.004]), and its MAE is 0.65~dBZ higher. The 90-min performance diagram (Fig.~\ref{fig:overall}(i)) further characterizes this difference: at 30 and 40~dBZ, MAG-Net maintains substantially higher POD than the learned baselines, whereas the latter strongly underpredict intense echoes, with POD below 0.1 at 40~dBZ. MAG-Net also has a higher success ratio than pySTEPS at these thresholds. pySTEPS is weaker than the other models on most metrics under the present protocol of four 10-min input frames and nine 10-min forecast leads.

Overall, the external comparison indicates that MAG-Net's main advantage lies in retaining high-reflectivity convective echoes rather than in uniformly improving all forecast metrics.

\begin{table}[!t]
  \centering
  \caption{OVERALL PERFORMANCE AND SATELLITE-CHANNEL ABLATION ON THE 2023 EVALUATION COHORT. VALUES ARE MEANS OVER NINE 10-MIN FORECAST LEADS. FSS$_{40}$ DENOTES 40-dBZ FSS AT A 5-km SCALE.}
  \label{tab:performance_ablation}
  \renewcommand{\arraystretch}{1.08}
  \setlength{\tabcolsep}{2.4pt}
  \footnotesize

  \textit{(a) External radar-only baselines}\par\smallskip
  \begin{tabularx}{\columnwidth}{@{}>{\raggedright\arraybackslash}Xrrrrr@{}}
    \toprule
    Model &
    CSI$_{20}\uparrow$ &
    CSI$_{30}\uparrow$ &
    CSI$_{40}\uparrow$ &
    FSS$_{40}\uparrow$ &
    RMSE$\downarrow$ \\
    \midrule
    pySTEPS (LK)          & 0.528 & 0.392 & 0.193 & 0.450 & 6.86 \\
    EarthFormer           & 0.634 & 0.472 & 0.158 & 0.355 & 4.90 \\
    FlowCast              & \textbf{0.638} & 0.493 & 0.189 & 0.416 & \textbf{4.89} \\
    SimVP-v2              & 0.617 & 0.467 & 0.160 & 0.355 & 5.19 \\
    \midrule
    MAG-Net               & 0.632 & \textbf{0.494} & \textbf{0.273} & \textbf{0.536} & 4.96 \\
    \bottomrule
  \end{tabularx}

  \vspace{0.55em}
  \textit{(b) Satellite-channel ablation}\par\smallskip
  \begin{tabularx}{\columnwidth}{@{}>{\raggedright\arraybackslash}Xrrrrr@{}}
    \toprule
    Configuration &
    CSI$_{20}\uparrow$ &
    CSI$_{30}\uparrow$ &
    CSI$_{40}\uparrow$ &
    FSS$_{40}\uparrow$ &
    RMSE$\downarrow$ \\
    \midrule
    Radar only (RD-Dual) & 0.629 & 0.482 & 0.252 & 0.512 & 5.06 \\
    Radar + IR & 0.630 & 0.488 & 0.265 & 0.527 & 4.99 \\
    Radar + IR + WV & 0.629 & 0.485 & \textbf{0.277} & \textbf{0.537} & 5.03 \\
    Radar + IR + WV + BTD$_{12-13}$ (MAG-Net) & 0.632 & 0.494 & 0.273 & 0.536 & \textbf{4.96} \\
    Five-band / nine-field AGRI & \textbf{0.637} & \textbf{0.498} & 0.274 & 0.534 & 4.99 \\
    \bottomrule
  \end{tabularx}

  \vspace{0.35em}
  \parbox{\columnwidth}{\raggedright\footnotesize All configurations in (b) use the matched Dual-Head/GPF formulation and are independently retrained.}
\end{table}

\subsection{Output Design, Fusion, and Satellite Channels}

\label{sec:res_ablation}

The dominant change in severe-echo skill arises from the output design. With radar input only, the regression output reaches CSI$_{40}=0.092$, whereas the corresponding Dual-Head/GPF output reaches 0.252. A similar contrast appears with satellite input: MAG-Net (regression head) reaches 0.127 and the independently trained Multimodal Regression model 0.120, compared with 0.273 for the full MAG-Net GPF output. The regression-based outputs retain the lowest pixel-wise errors, with RMSE values of 4.87--4.89~dBZ. An independently trained classification-only model reaches a slightly higher CSI$_{40}$ (0.281 versus 0.273), but its output consists of threshold-exceedance probabilities rather than a continuous reflectivity field. GPF therefore recovers most of the high-threshold classification skill while retaining a continuous reflectivity prediction, with a moderate increase in RMSE relative to the regression output (4.96 versus 4.89~dBZ).

After controlling for the output formulation, the remaining satellite contribution is smaller but statistically supported. Under the matched Dual-Head/GPF configuration, adding the three AGRI fields increases CSI$_{40}$ from 0.252 to 0.273, corresponding to a paired difference of $+0.021$ [0.018, 0.024]. The 40-dBZ FSS increases by $+0.024$ [0.020, 0.028], while RMSE and MAE (not shown) decrease by 0.09 and 0.11~dBZ, respectively.

The channel ablation in Table~\ref{tab:performance_ablation}(b) further shows that satellite benefit is not monotonic in the number of input fields. Radar+IR produces a smaller CSI$_{40}$ increment than the default three-field configuration, while Radar+IR+WV gives the largest CSI$_{40}$ gain but little improvement at 20 and 30~dBZ. The five-band, nine-field variant produces larger gains at lower reflectivity thresholds but no consistent additional benefit at 40~dBZ. Thus, different channel combinations emphasize different parts of the forecast distribution, and no tested configuration dominates across all metrics. The default three-field configuration is retained as a compact setting that provides a statistically supported improvement over the matched RD-Dual model rather than as an empirically optimal channel set.

\subsection{Qualitative Example}

\label{sec:res_case}

\begin{figure*}[!t]

  \centering

  \includegraphics[width=\textwidth]{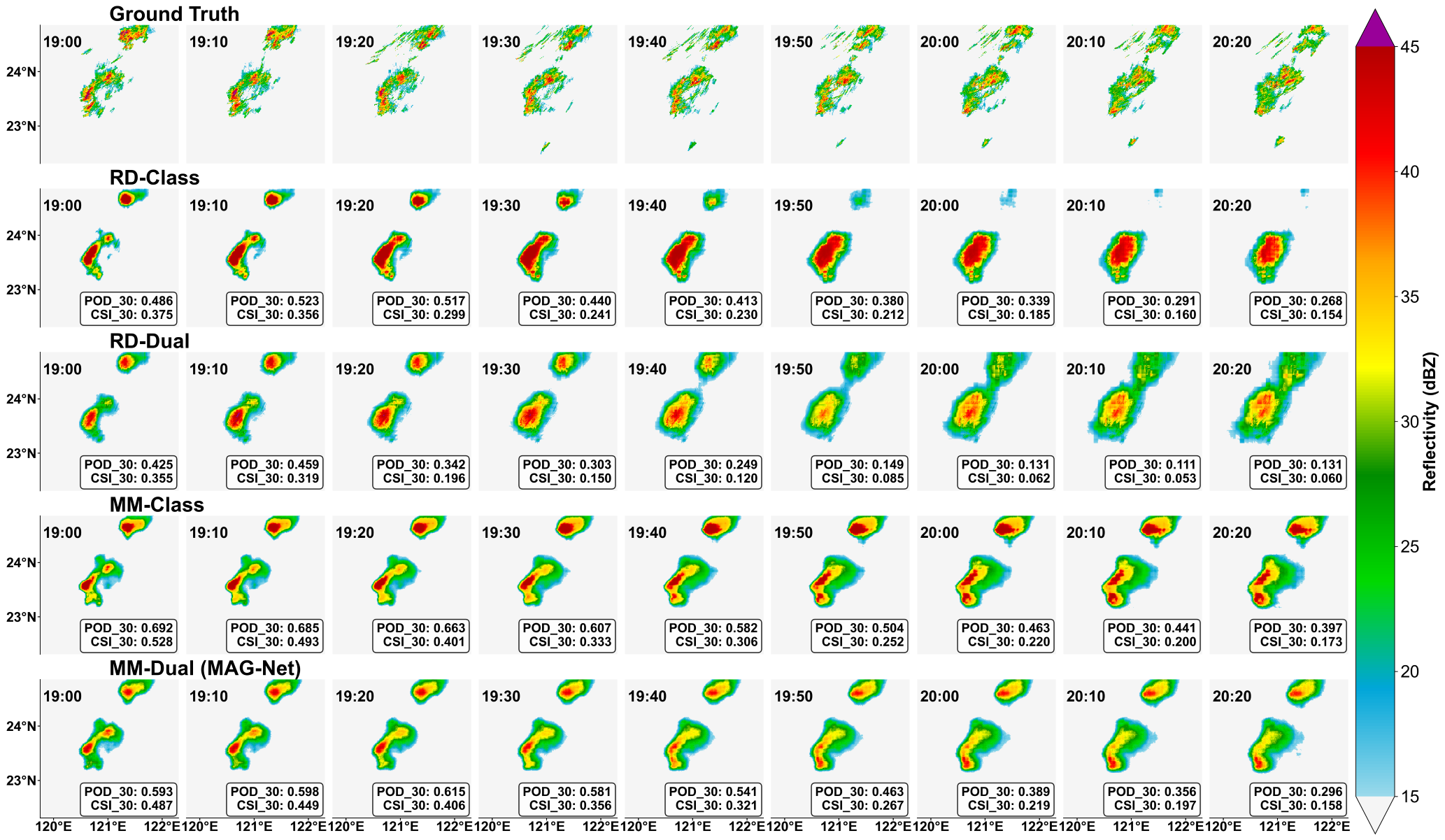}

  \caption{Representative severe-echo evolution case from 2023. Rows show observed composite reflectivity and 10--90-min forecasts from the radar-only classification (RD-Class), radar-only Dual-Head (RD-Dual), multimodal classification (MM-Class), and multimodal Dual-Head (MAG-Net) variants. Dual-Head rows show the GPF output. POD and CSI at 30~dBZ are reported for each forecast panel. This case qualitatively illustrates forecast behavior consistent with the cohort-level evaluation and is not used to quantify the satellite increment.}

  \label{fig:case}

\end{figure*}

Fig.~\ref{fig:case} illustrates the forecast behavior identified in the cohort-level evaluation. As lead time increases, the radar-only Dual-Head forecast progressively weakens and broadens the intense echo core, whereas the multimodal forecasts retain a more compact high-reflectivity structure. The corresponding CSI$_{30}$ also decreases more slowly for the multimodal variants. This example is intended only as a qualitative illustration. The satellite increment is quantified by the population-level analyses in Table~\ref{tab:performance_ablation} and the following sections.

\subsection{Lifecycle-Stratified Skill}

\label{sec:res_lifecycle}

\begin{figure}[!t]
  \centering
  \includegraphics[width=\linewidth]{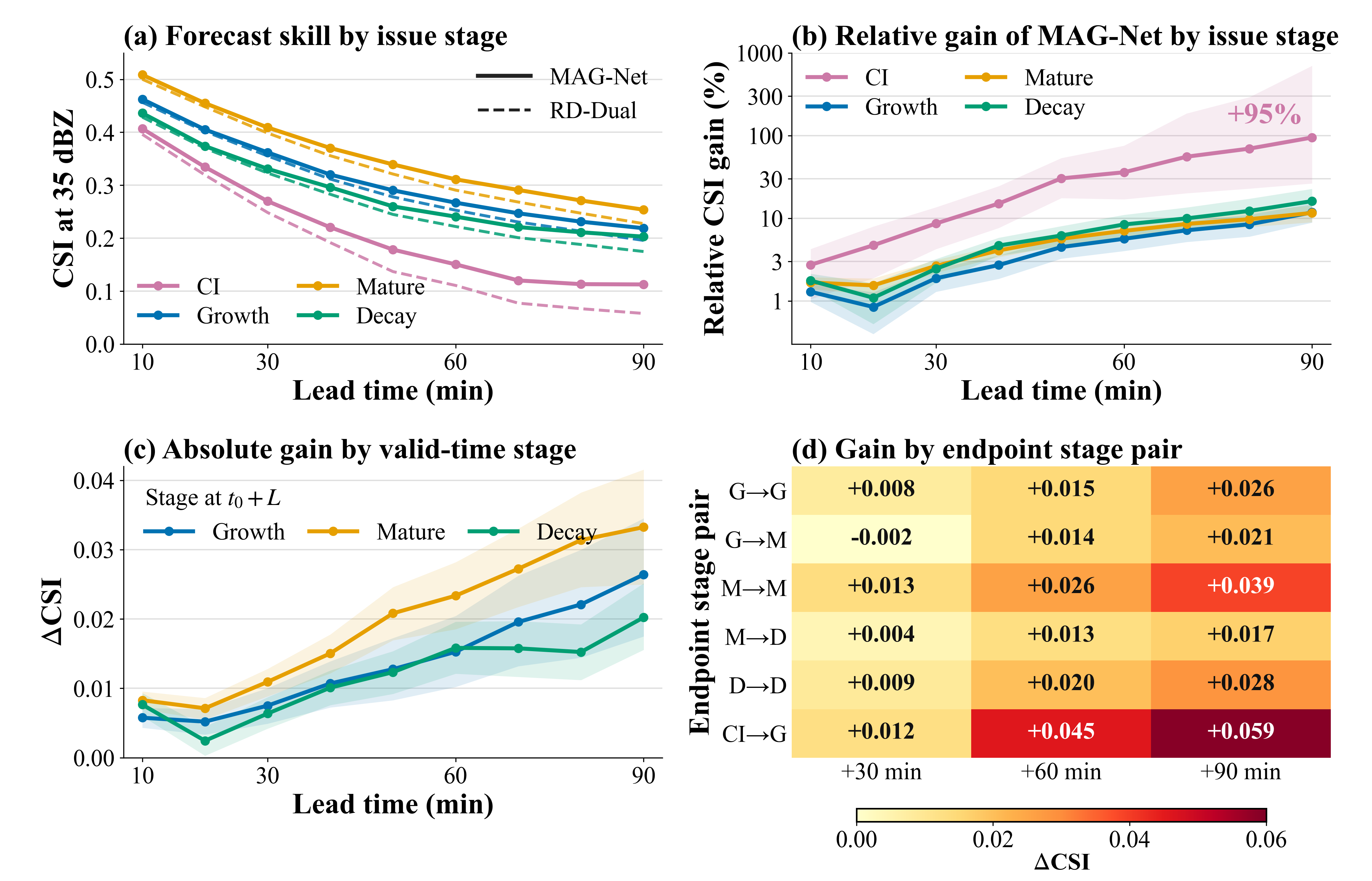}
  \caption{Lifecycle-stratified forecast skill of MAG-Net and the matched RD-Dual model, evaluated using CSI at 35~dBZ in object-following windows (8,735 issues from 6,361 tracks). Gains in (b)--(d) are MAG-Net minus RD-Dual. (a) CSI by issue-time lifecycle stage. Solid lines denote MAG-Net and dashed lines RD-Dual. (b) Relative CSI gain by issue stage. Shading denotes 95\% confidence intervals from 2,000 paired bootstrap resamples of tracked objects. (c) Absolute CSI gain grouped by lifecycle stage at valid time $t_0+L$. (d) Absolute gain by issue-to-valid stage transition $S(t_0)\rightarrow S(t_0+L)$ at 30, 60, and 90~min. Sample counts and corresponding confidence intervals are provided in Supplementary Table~S2.}
  \label{fig:lifecycle}
\end{figure}

Fig.~\ref{fig:lifecycle} and Supplementary Table~S2 show that the satellite increment varies with convective lifecycle and forecast lead time. Absolute CSI$_{35}$ is highest for forecasts initialized during the Mature stage and lowest for CI at all leads (Fig.~\ref{fig:lifecycle}(a)). Nevertheless, MAG-Net outperforms the matched RD-Dual model across all issue-time stages, and the separation generally increases toward longer lead times.

At 90~min, the paired CSI$_{35}$ increments are $+0.023$ [0.017, 0.029] for Growth, $+0.026$ [0.020, 0.032] for Mature, and $+0.028$ [0.020, 0.036] for Decay. CI shows a larger increment of $+0.055$ [0.014, 0.097]. Relative to an RD-Dual CSI of 0.058, this is a gain of approximately 95\%, compared with about 10--15\% for the other stages (Fig.~\ref{fig:lifecycle}(b)). The CI cohort contains 201 tracks, but the 90-min estimate uses 189 issues from 111 of them and consequently has substantially wider uncertainty than the other stage estimates.

Stratification by the observed stage at valid time provides a complementary view (Fig.~\ref{fig:lifecycle}(c)). The largest absolute satellite increment occurs when the verifying target is in the Mature stage, indicating that the benefit is not restricted to forecasts initialized before convective development. Endpoint-pair analysis further distinguishes stage persistence from lifecycle evolution (Fig.~\ref{fig:lifecycle}(d)). At 60 and 90~min, positive and statistically supported increments occur for both persistent trajectories, such as Mature$\rightarrow$Mature, and evolving trajectories, including Growth$\rightarrow$Mature and Mature$\rightarrow$Decay. For example, Mature$\rightarrow$Mature reaches an increment of $+0.039$ at 90~min. CI$\rightarrow$Growth has the largest point estimates, $+0.045$ at 60~min and $+0.059$ at 90~min, although the latter is based on fewer than 100 issues. At 30~min, the Growth$\rightarrow$Mature difference is not statistically supported.

Taken together, the lifecycle analysis shows that the satellite contribution is not confined to convective initiation. It is detectable across growth, mature, and decay stages, including forecasts that undergo a stage transition, while its relative importance is greatest for CI, where the RD-Dual baseline is weakest.

\subsection{Robustness to Input Misalignment}

\label{sec:res_coreg}

Supplementary Fig.~S1 and Table~S1 evaluate the sensitivity of the fixed MAG-Net checkpoint to perturbations of the satellite input. Residual spatial changes at approximately one native AGRI pixel, including one-pixel shifts and nearest-neighbor rather than bilinear resampling, alter CSI$_{40}$ by less than 0.001. Larger spatial and temporal mismatches produce progressively larger degradation: 4- and 8-pixel shifts reduce CSI$_{40}$ by approximately 0.003 and 0.008, respectively, while a $\pm60$~min temporal offset reduces it by 0.012--0.014. Pairing radar with satellite observations from a different outbreak causes a much larger reduction of 0.093, taking performance below the matched RD-Dual model.

These perturbation tests indicate that the measured satellite increment depends on correctly matched satellite context, while being relatively insensitive to small residual co-registration errors.

\subsection{Observation-Conditioned Satellite Increment}

\label{sec:res_products}

\begin{figure}[!t]
  \centering
  \includegraphics[width=\linewidth]{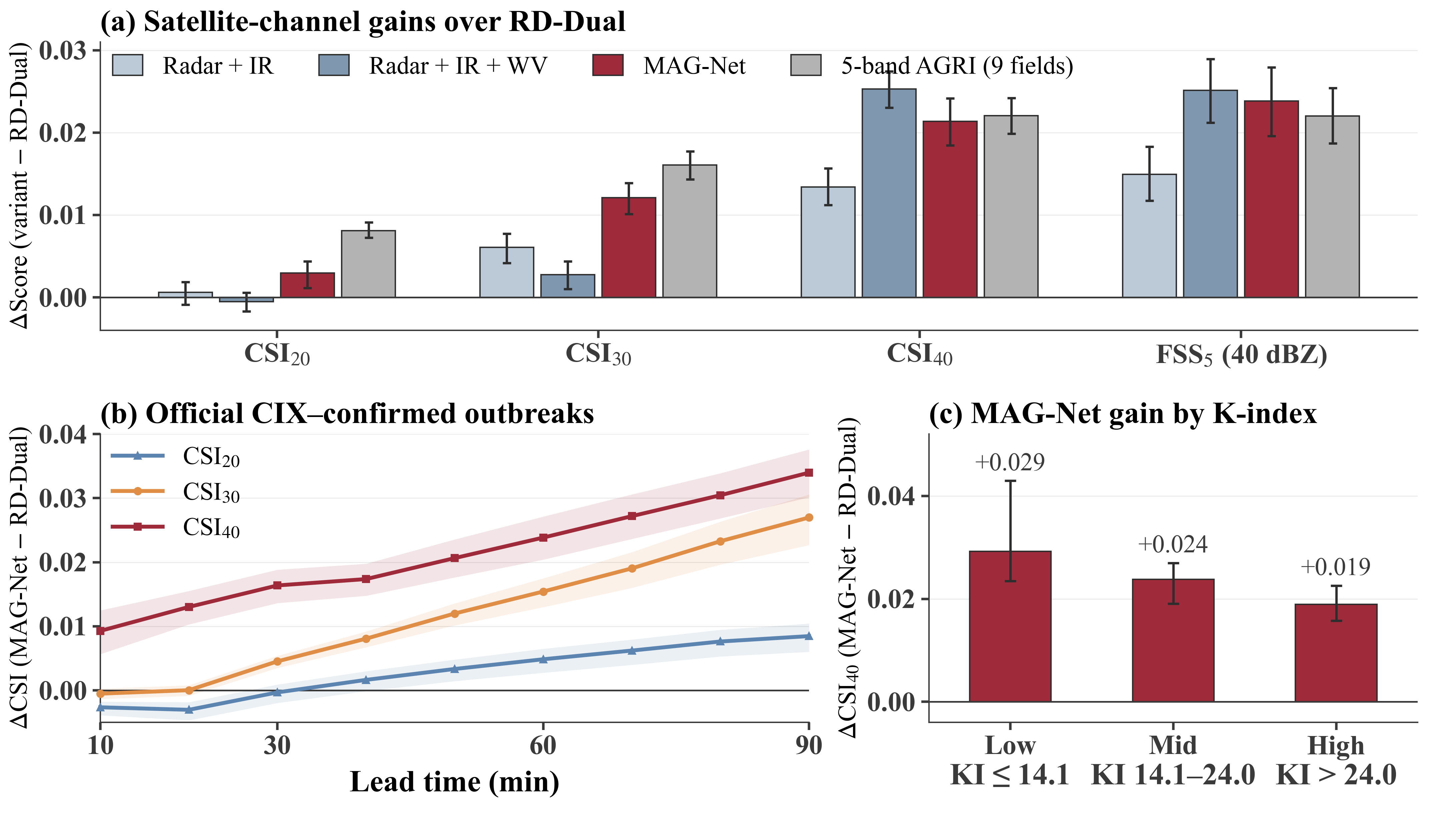}
  \caption{Satellite-channel variants and observation-conditioned analyses. (a) Paired gains of independently retrained satellite-input variants over the matched RD-Dual model for CSI at 20, 30, and 40~dBZ and 40-dBZ FSS at 5~km. RD-Dual reference scores are 0.629, 0.482, 0.252, and 0.512, respectively. (b) MAG-Net $-$ RD-Dual CSI gains at 20, 30, and 40~dBZ versus forecast lead time for the 105 outbreaks containing an operational AGRI CIX detection at or before initialization that was subsequently associated with observed radar reflectivity $\geq30$~dBZ within 20~km (73,350 sequences). (c) CSI$_{40}$ gain stratified by tertiles of the outbreak-median FY-4B GIIRS AII K-index (DQF$\leq1$). Error bars and shaded bands denote outbreak-cluster bootstrap 95\% confidence intervals. CIX and AII are used only for post-hoc stratification and are not model inputs.}
  \label{fig:physical}
\end{figure}

Fig.~\ref{fig:physical}(a) shows the channel comparison summarized in Table~\ref{tab:performance_ablation}(b). The satellite increment remains positive within the official-product-conditioned subsets in Fig.~\ref{fig:physical}(b) and (c). Among the 105 outbreaks containing radar-confirmed AGRI CIX signals, the CSI$_{40}$ increment increases from $+0.009$ [0.006, 0.013] at 10~min to $+0.034$ [0.030, 0.038] at 90~min (Fig.~\ref{fig:physical}(b)). Because this subset contains 99\% of the evaluated sequences, the result primarily confirms that the lead-time-dependent satellite increment is also present in outbreaks associated with operational CIX signals.

Stratification by the FY-4B GIIRS K-index shows positive CSI$_{40}$ increments across all three instability tertiles (Fig.~\ref{fig:physical}(c)). The increments are $+0.029$ [0.023, 0.043], $+0.024$ [0.019, 0.027], and $+0.019$ [0.016, 0.023] from the low- to high-K-index groups, respectively. The high-minus-low interaction is $-0.010$ [$-0.024$, $-0.003$]. Thus, the satellite benefit is not confined to a particular instability regime and does not increase with the K-index in this cohort.

\subsection{Integrated-Gradients Diagnostics}

\label{sec:res_ig}

\begin{figure}[!t]
  \centering
  \includegraphics[width=\linewidth]{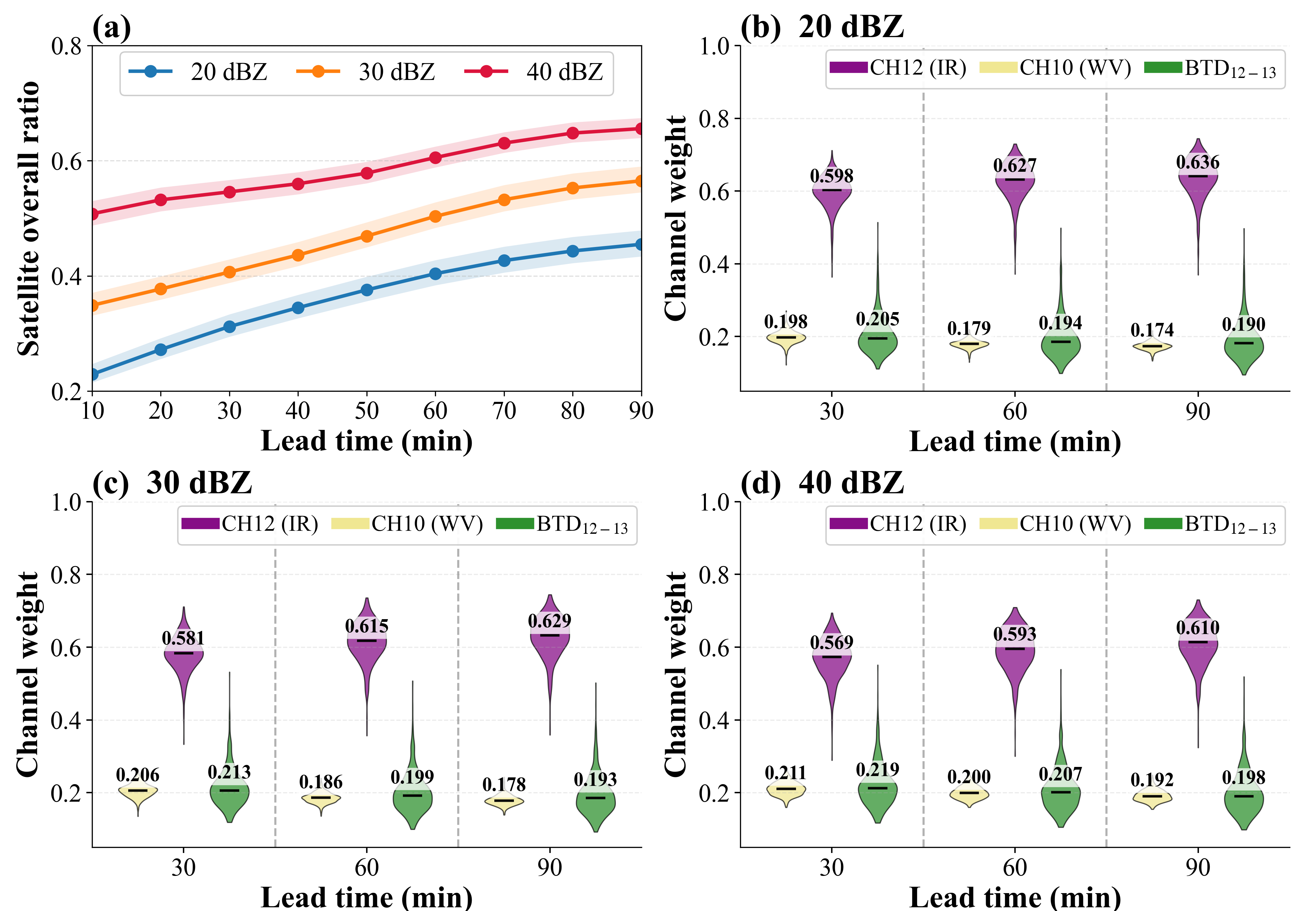}
  \caption{Integrated-Gradients input-sensitivity diagnostics for the MAG-Net checkpoint (500 sequences from 72 outbreaks, using a zero baseline in normalized space and a 50-step midpoint approximation). The attribution target is the summed exceedance probability at 20, 30, or 40~dBZ. (a) Satellite share of mean absolute attribution versus forecast lead time. Shaded bands denote outbreak-cluster bootstrap 95\% confidence intervals. (b)--(d) Per-sequence fraction of satellite attribution assigned to CH12, CH10, and BTD$_{12-13}$ at 30, 60, and 90~min for the 20-, 30-, and 40-dBZ targets, respectively. Values denote means and horizontal bars medians. Because attribution magnitude depends on the chosen baseline, these diagnostics are interpreted as model sensitivity rather than as unique information content or evidence of a physical mechanism.}
  \label{fig:ig}
\end{figure}

Integrated Gradients shows a consistent increase in model sensitivity to the satellite inputs with forecast lead time (Fig.~\ref{fig:ig}(a)). The satellite share of absolute attribution rises from 0.23 to 0.46 for the 20-dBZ target, from 0.35 to 0.57 for 30~dBZ, and from 0.51 [0.49, 0.53] to 0.66 [0.64, 0.67] for 40~dBZ between 10 and 90~min. The increase occurs for the large majority of evaluated sequences and is statistically supported for all three thresholds. At each lead, the satellite attribution share is also larger for higher-reflectivity targets. For example, the paired difference between the 40- and 20-dBZ targets is $+0.28$ [0.26, 0.29] at 10~min and $+0.20$ [0.19, 0.22] at 90~min.

Within the satellite inputs, CH12 consistently receives the largest fraction of attribution (Fig.~\ref{fig:ig}(b)--(d)). Its mean share increases modestly with lead time, whereas CH10 and BTD$_{12-13}$ each contribute smaller fractions that decline slightly. The numerical IG approximation is well converged, with median relative completeness errors below 1.6\% across all thresholds and leads.

The absolute attribution fractions depend on the choice of baseline and should not be interpreted as the relative information content of radar and satellite observations. In particular, the zero baseline in normalized satellite space corresponds to 200~K for CH10 and CH12 rather than to a neutral cloud-free state. We therefore interpret the IG results only in terms of relative trends under the same baseline: the increase with forecast lead, the ordering across reflectivity thresholds, and the relative sensitivity among satellite channels. These patterns are consistent with the matched experiments, in which the satellite increment becomes larger at longer lead times, but they remain sensitivity diagnostics and do not establish how the model uses the satellite information or whether it has learned a physical mechanism.

\subsection{Seasonal and Spatial Robustness}

\label{sec:res_season}

\begin{figure}[!t]
  \centering
  \includegraphics[width=\linewidth]{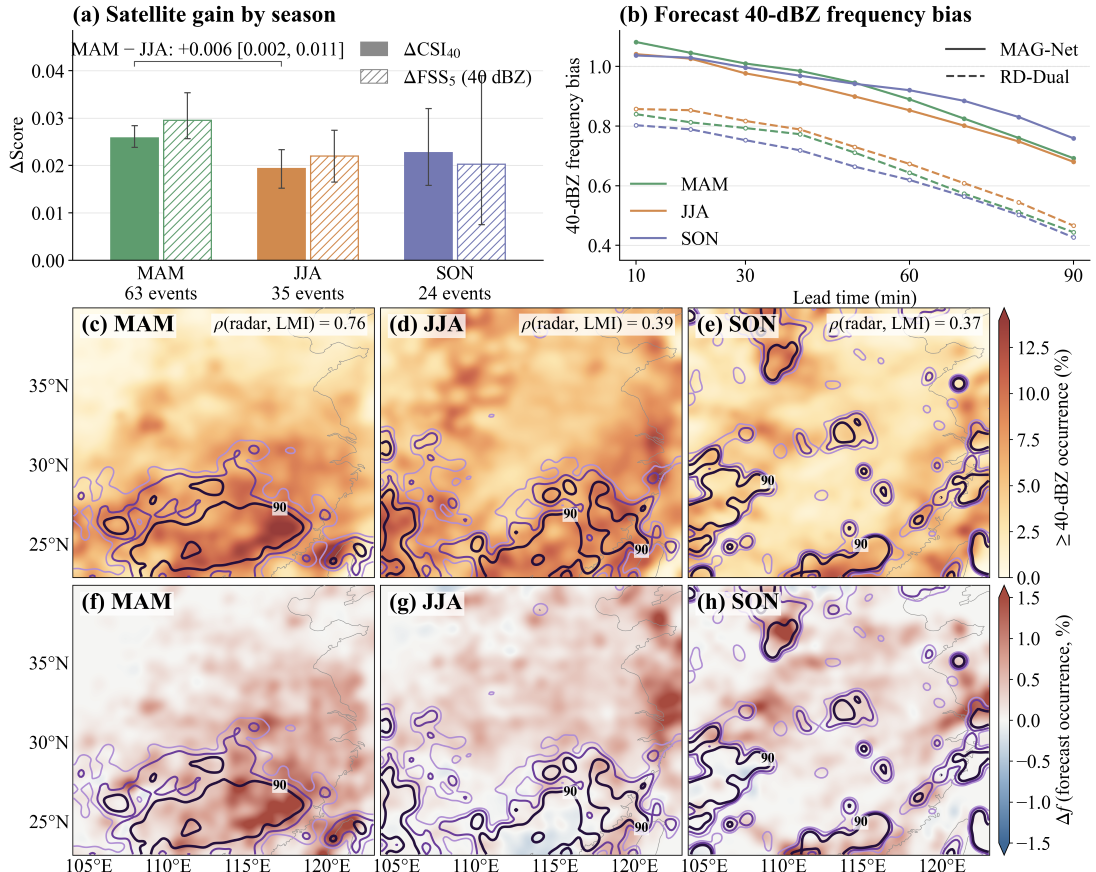}
  \caption{Seasonal and spatial robustness within the study region. Gains in (a) and (f)--(h) are MAG-Net minus RD-Dual. (a) CSI$_{40}$ and 40-dBZ FSS at 5~km by season, with outbreak-cluster bootstrap 95\% confidence intervals. The MAM$-$JJA difference is also shown. (b) Seasonal 40-dBZ frequency bias of MAG-Net and RD-Dual versus forecast lead time. (c)--(e) Observed occurrence frequency of $\geq$40-dBZ echoes at $T+60$~min for MAM, JJA, and SON. The symbol $\rho$ denotes the spatial Spearman correlation between radar occurrence and FY-4 LMI flash density. (f)--(h) MAG-Net $-$ RD-Dual difference in forecast $\geq$40-dBZ occurrence frequency at $T+60$~min. LMI contours denote the seasonal 70th, 80th, and 90th percentiles of flash density. Maps characterize the 2023 SCE-R evaluation cohort and should not be interpreted as a climatology.}
  \label{fig:season}
\end{figure}

The increment of MAG-Net over RD-Dual remains positive across the three represented seasons (Fig.~\ref{fig:season}(a)). For CSI$_{40}$, the paired difference is statistically supported in spring (MAM, 63 outbreaks), summer (JJA, 35), and autumn (SON, 24).

The 40-dBZ frequency bias provides additional context for this result (Fig.~\ref{fig:season}(b)). RD-Dual increasingly underpredicts intense echoes with forecast lead time, reaching a bias of approximately 0.45 at 90~min. MAG-Net also loses high-reflectivity occurrence with lead time, but less rapidly, decreasing from slightly above unity at short leads to approximately 0.70--0.75 at 90~min. The satellite input therefore reduces, but does not eliminate, the progressive underprediction of intense echoes.

The spatial analyses in Fig.~\ref{fig:season}(c)--(h) characterize the 2023 SCE-R evaluation cohort rather than a regional climatology. Regions with frequent observed 40-dBZ echoes broadly overlap regions of enhanced LMI lightning density, although their spatial Spearman rank correlation varies by season (0.76 in MAM, 0.39 in JJA, and 0.37 in SON). Relative to the RD-Dual model, MAG-Net produces more frequent 40-dBZ echoes over much of the study domain. Its spatial agreement with the observed occurrence field is generally comparable to the RD-Dual forecast, with correlations of 0.86 versus 0.84 in MAM, 0.44 versus 0.46 in JJA, and 0.58 versus 0.51 in SON.

Together, these results show that the satellite increment is not restricted to a single season or localized subset of the 2023 study domain. They provide within-region robustness and spatial consistency checks, but do not test geographic generalization beyond southeastern China.

\section{Discussion}

\label{sec:discussion}

\subsection{Interpreting the Satellite Contribution}

The controlled experiments indicate that the satellite contribution is smaller than the effect of the output formulation, but is systematic rather than incidental. The increment increases toward longer lead times, is reduced by large spatial or temporal mismatches, and disappears when radar and satellite observations are paired across different outbreaks. It is also detectable across the observed convective lifecycle rather than being confined to a single stage. In relative terms, the largest benefit occurs before the birth of a persistent radar core, where RD-Dual skill is lowest, whereas the largest absolute gains occur for forecasts verifying during the mature stage. Together, these results suggest that the value of satellite observations increases when the recent radar state becomes less sufficient to describe subsequent convective evolution.

This interpretation is consistent with the observational complementarity of the AGRI inputs. CH12 provides information on cloud-top temperature and its evolution, while CH10 provides mid-tropospheric water-vapor context. Such information can evolve differently from the radar-observed precipitation field before and during changes in convective intensity. However, CH10 alone is not a measure of atmospheric stability, and the radiance-derived BTD$_{12-13}$ should not be assigned a unique physical interpretation. Consistent with this limitation, the satellite increment remains positive across GIIRS K-index strata but does not increase with increasing K-index. The Integrated-Gradients results likewise show increasing model sensitivity to the satellite inputs at longer leads, particularly to CH12, but sensitivity does not establish unique information content or a learned physical mechanism.

Satellite information contributes during initiation, growth, mature, and decay stages, including forecasts that cross lifecycle boundaries. Its contribution is thus state dependent but not stage specific. In the present dataset, the additional modality appears most valuable when extrapolation of the radar-observed state alone provides an incomplete description of subsequent echo evolution.

\subsection{Output Formulation and Multimodal Information}

A central result of this study is that improvements associated with multimodal nowcasting should be separated from improvements caused by the forecast-output formulation. In the matched radar-only experiments, replacing the regression output with the Dual-Head/GPF formulation increases CSI$_{40}$ from 0.092 to 0.252, whereas adding the three AGRI fields under the same Dual-Head/GPF configuration increases it further from 0.252 to 0.273. Thus, the output formulation and the satellite observations contribute jointly, and the larger share of the severe-echo improvement comes from how the forecast is represented and trained.

This distinction is important because regression and threshold-based objectives emphasize different aspects of the forecast distribution. The regression outputs achieve lower RMSE and MAE but substantially attenuate high-reflectivity echoes, whereas the classification output retains stronger threshold-defined structures but does not directly provide a continuous reflectivity field. GPF combines these two outputs and recovers most of the high-threshold classification skill while maintaining a continuous forecast. It should therefore be interpreted as an operating-point choice rather than as a uniformly superior forecast representation: the improvement in severe-echo retention is accompanied by some degradation in pixel-wise error and lower-threshold skill.

The channel experiments lead to a similar conclusion for the multimodal component. Adding more satellite fields does not produce a monotonic improvement. The benefit therefore cannot be attributed simply to greater input dimensionality. Instead, the matched experiments support a narrower conclusion: correctly paired satellite observations provide additional predictive information under the chosen forecast formulation, but the magnitude and location of that benefit depend on the forecast target and channel combination.

\subsection{Scope, Transferability, and Limitations}

The present conclusions are intentionally restricted to the 90-min forecast horizon and to the southeastern-China CIDS observing system. Radar reflectivity retains substantial Lagrangian autocorrelation over approximately the first 1--2~h~\cite{germann2002scale}. This is therefore a regime in which radar extrapolation remains informative while errors associated with growth, decay, and structural evolution become increasingly important. The experiment was designed to measure the incremental value of satellite information within this radar-dominated regime rather than to determine whether satellite observations become a primary predictor at longer lead times.

The seasonal analysis provides evidence of robustness within the study region, but not geographic generalization. As a bounded transfer experiment, the RD-Dual formulation was additionally trained and evaluated on a held-out event split of the Storm EVent ImagRy (SEVIR) dataset~\cite{veillette2020sevir} (Supplementary Fig.~S2). SEVIR differs substantially from CIDS in observing system, temporal resolution, target variable, and forecast protocol, using vertically integrated liquid at 5-min intervals with a $13\rightarrow12$ sequence configuration. The persistence of a Dual-Head/GPF benefit in this setting suggests that the output-design effect is not specific to the CIDS radar data. Because SEVIR does not provide satellite channels equivalent to the AGRI CH10, CH12, and BTD$_{12-13}$ triplet, this experiment provides no evidence that the multimodal satellite contribution transfers across regions, sensors, or climates. A multimodal experiment on SEVIR would require a different satellite input configuration and would not be a like-for-like transfer test.

Several limitations should therefore be considered when interpreting the reported skill. First, the AGRI observations retain their approximately 4-km sampling scale after resampling and cannot provide radar-scale spatial detail. Second, the lifecycle labels are defined retrospectively and depend on both the tracking procedure and the 35-dBZ core definition. The CI subset is also substantially smaller than the other lifecycle strata. Third, the CIX analysis is conditioned on an AGRI-derived product and is therefore a consistency check rather than independent validation. Finally, the external baselines were retrained under a common protocol that differs from some of their original configurations. 

\subsection{Computational Implications}

MAG-Net remains non-autoregressive, producing all nine forecast leads in a single forward pass. The full model contains 336~M parameters and requires 184~GFLOPs per $256\times256$ sample, compared with 151~GFLOPs for the matched RD-Dual model. The satellite stream and fusion module account for only about 1.1~M additional parameters, and the inference-time GPF operation introduces no learned parameters.

On a single NVIDIA H100 GPU in FP32, the additional satellite pathway increases end-to-end inference time by approximately 4.7~ms per sample relative to the RD-Dual configuration, while GPF itself adds less than 1~ms. These measurements exclude data loading and decoding and are therefore intended as model-level computational references rather than estimates of operational latency. Detailed timing and memory measurements are provided for reproducibility rather than as evidence of operational readiness.

\section{Conclusion}

\label{sec:conclusion}

This study separates two effects that are often conflated in multimodal convective nowcasting: improvements arising from the forecast-output formulation and those attributable to additional satellite observations. In the matched experiments, the Dual-Head/GPF formulation and the FY-4A AGRI observations contribute jointly to severe-echo skill. The larger share comes from the output formulation, and the AGRI fields add a further, statistically supported increment beyond the corresponding RD-Dual model.

The satellite contribution is not confined to convective initiation or to a single environmental regime. It becomes more pronounced at longer forecast leads and is detectable across growth, mature, and decay stages, including forecasts that undergo lifecycle transitions. Its relative benefit is largest before radar-core birth, where RD-Dual skill is weakest, whereas larger absolute gains also occur for mature targets. Together, these results indicate that the value of satellite observations is state dependent: additional satellite-observed cloud and moisture context becomes more useful when the recent radar-observed state alone is less sufficient to describe subsequent convective evolution.

More broadly, the results suggest that multimodal nowcasting should be evaluated not only by whether an additional modality improves aggregate forecast skill, but also by separating that improvement from architectural and output-design effects and identifying when the additional observations provide complementary predictive information. Within the present study region, the FY-4A satellite fields satisfy this narrower criterion. The analyses do not establish that MAG-Net has learned specific convective physical mechanisms, nor do they demonstrate transfer of the multimodal benefit beyond the southeastern-China observing system.

\section*{Data Availability Statement}
The CIDS repository and reading code, including the public intensity-level radar representation and satellite features, are available from the Science Data Bank at \url{https://doi.org/10.57760/sciencedb.30286}~\cite{liu2026cids}. The 2023 products used only for post-hoc checks were obtained from the NSMC Fengyun satellite data portal (\url{https://satellite.nsmc.org.cn/DataPortal/}): the FY-4A AGRI convective-initiation product (CIX), the FY-4B GIIRS Atmosphere Instability Index (AII; K-index), and FY-4A LMI lightning events (LMIE). The operational CMA composite reflectivity fields used for dBZ evaluation (\url{http://data.cma.cn/data/cdcdetail/dataCode/J.0019.0010.S001.html}) are subject to CMA data policies. Requests should be directed to CMA through the appropriate data-access procedures.

\section*{Acknowledgment}
The authors gratefully acknowledge the China Meteorological Administration (CMA) and the National Satellite Meteorological Center for providing the radar and satellite observations used in this study. We also thank the open-source community for making their code publicly available, which greatly facilitated the comparative experiments. Specifically, we acknowledge the official implementation of CPrecNet provided by Park and Lee via Zenodo~\cite{park2024cprecnet_code}, SimVP-v2 within the OpenSTL framework~\cite{tan2023openstl}, and the public implementations of pySTEPS, EarthFormer, and FlowCast.

\bibliographystyle{IEEEtran}
\bibliography{refs}

\end{document}